%% file: main.tex
\documentclass[sigconf,screen]{acmart}
\pdfoutput=1

\usepackage[export]{adjustbox}
\usepackage[normalem]{ulem}
\usepackage{multirow}
\usepackage{pifont}
\usepackage{listings}
\usepackage{mathpartir}
\usepackage{xspace}
\usepackage{tikz}
\usepackage{graphicx}
\usepackage{siunitx}
\usepackage{mdframed}
\usepackage{mathtools}
\usepackage[nounderscore]{syntax}
\usepackage{subcaption}
\usepackage[linesnumbered,ruled,vlined]{algorithm2e}
\usepackage{flushend}
\usepackage{mathtools}
\usepackage{mathpartir}
\usepackage{amsthm}
\usepackage[export]{adjustbox}
\usepackage{xcolor,colortbl}
\usepackage{multirow}
\usepackage{amsfonts}
\usepackage{booktabs}

\newtheorem{definition}{Definition}[section]

\newcommand{\rulesep}{\unskip\ \vrule\ }
\lstdefinestyle{customc}{
  belowcaptionskip=1\baselineskip,
  breaklines=true,
  numbers=left,
  xleftmargin=2.5em,
  language=C,
  showstringspaces=false,
  basicstyle=\footnotesize\ttfamily,
  keywordstyle=\bfseries\color{black},
  commentstyle=\itshape\color{gray},
  identifierstyle=\color{black},
  stringstyle=\color{orange},
  moredelim=**[is][\color{red}]{@}{@},
  moredelim=**[is][\color{green!40!black}]{@@}{@@},
}

\SetKwProg{Func}{Function}{}{}
\SetAlFnt{\small\rmfamily}
\SetKw{Continue}{continue}
\SetKw{Break}{break}
\SetKw{Case}{case}
\SetKw{And}{and}
\SetKw{Or}{or}
\SetKw{Not}{not}
\SetKw{In}{in}
\SetKw{Is}{is}
\SetKw{Null}{NULL}
\SetKwInput{Input}{Input}
\SetKwInOut{Output}{Output}

\AtBeginDocument{%
  \providecommand\BibTeX{{%
    \normalfont B\kern-0.5em{\scshape i\kern-0.25em b}\kern-0.8em\TeX}}}
\setcopyright{none}

\setcopyright{rightsretained}
\acmPrice{}
\acmDOI{10.1145/3611643.3616301}
\acmYear{2023}
\copyrightyear{2023}
\acmSubmissionID{fse23main-p539-p}
\acmISBN{979-8-4007-0327-0/23/12}
\acmConference[ESEC/FSE '23]{Proceedings of the 31st ACM Joint European Software Engineering Conference and Symposium on the Foundations of Software Engineering}{December 3--9, 2023}{San Francisco, CA, USA}
\acmBooktitle{Proceedings of the 31st ACM Joint European Software Engineering Conference and Symposium on the Foundations of Software Engineering (ESEC/FSE '23), December 3--9, 2023, San Francisco, CA, USA}
\received{2023-02-02}
\received[accepted]{2023-07-27}

\newcommand{\toolname}{{\sc PEM}\xspace}

\newcommand{\revise}[1]{{\color{black}{#1}}}
\newcommand{\rebuttal}[1]{{\color{black}{#1}}}
\newcommand{\rebuttalshort}[1]{{\color{black}{#1}}}
\definecolor{Gray}{gray}{0.935}
\newcolumntype{g}{>{\columncolor{Gray}}c}
\newcolumntype{G}{>{\columncolor{Gray}}r}

\begin{document}

\title{\toolname: Representing Binary Program Semantics for Similarity Analysis via a Probabilistic Execution Model}

\author{Xiangzhe Xu}
\email{xu1415@purdue.edu}
\authornote{Both authors contributed equally to this research.}
\author{Zhou Xuan}
\authornotemark[1]
\email{xuan1@purdue.edu}
\affiliation{%
  \institution{Purdue University}  
  \city{West Lafayette}
  \country{USA}  
}

\author{Shiwei Feng}
\affiliation{%
  \institution{Purdue University}  
  \city{West Lafayette}
  \country{USA}  
}
\email{feng292@purdue.edu}

\author{Siyuan Cheng}
\affiliation{
  \institution{Purdue University}  
  \city{West Lafayette}
  \country{USA}  
}
\email{cheng535@purdue.edu}

\author{Yapeng Ye}
\affiliation{%
  \institution{Purdue University}  
  \city{West Lafayette}
  \country{USA}  
}
\email{ye203@purdue.edu}

\author{Qingkai Shi}
\affiliation{
  \institution{Purdue University}  
  \city{West Lafayette}
  \country{USA}  
}
\email{shi553@purdue.edu}

\author{Guanhong Tao}
\affiliation{
  \institution{Purdue University}  
  \city{West Lafayette}
  \country{USA}  
}
\email{taog@purdue.edu}

\author{Le Yu}
\affiliation{
  \institution{Purdue University} 
  \city{West Lafayette}
  \country{USA}  
}
\email{yu759@purdue.edu}

\author{Zhuo Zhang}
\affiliation{
  \institution{Purdue University}  
  \city{West Lafayette}
  \country{USA}  
}
\email{zhan3299@purdue.edu}

\author{Xiangyu Zhang}
\affiliation{
  \institution{Purdue University}  
  \city{West Lafayette}
  \country{USA}  
}
\email{xyzhang@cs.purdue.edu}

\renewcommand{\shortauthors}{X. Xu, Z. Xuan, S. Feng, S. Cheng, Y. Ye, G. Tao, Q. Shi, Z. Zhang, and X. Zhang}

\begin{abstract}
  \input{abstract}

\end{abstract}



\maketitle
\thispagestyle{plain}
\pagestyle{plain}

\input{intro}

\input{background}

\input{method}

\input{eval}

\input{related}
\vspace{-5pt}
\section{Conclusion}
\vspace{-3pt}
We develop a novel probabilistic execution model for effective sampling and representation of binary program semantics. 
It features a path-sampling algorithm that is resilient to code transformations and a probabilistic memory model that can tolerate invalid memory accesses. It substantially outperforms the state of the arts. 

\vspace{-5pt}
\section{Data Availability}
\vspace{-3pt}

Our experimental data and the artifact are available at~\cite{PEM}.

\begin{acks}
We thank the anonymous reviewers for their valuable
comments and suggestions. This research was supported, in part by DARPA VSPELLS - HR001120S0058, IARPA TrojAI W911NF-19-S-0012, NSF 1901242 and 1910300, ONR N000141712045, N000141410468 and N000141712947. Any opinions, findings, and conclusions in this paper are those of the authors only and do not necessarily reflect the views of our sponsors.
\end{acks}

\bibliographystyle{ACM-Reference-Format}
\bibliography{prog-repr}

\clearpage

\appendix
\input{suppl}

\end{document}

%% file: abstract.tex
Binary similarity analysis determines if two binary executables are from the same source program. 
Existing techniques leverage static and dynamic program features and may utilize advanced Deep Learning techniques. Although they have demonstrated great potential, the community believes that a more effective representation of program semantics can further improve similarity analysis. In this paper, we propose a new method to represent binary program semantics. It is based on a novel probabilistic execution engine that can effectively sample the input space and the program path space of subject binaries. More importantly, it ensures that the collected samples are comparable across binaries, addressing the substantial variations of input specifications. 
Our evaluation on 9 real-world projects with 35k functions, and comparison with 6 state-of-the-art techniques show that \revise{\toolname can achieve a precision of 96\% with common settings, outperforming the baselines by 10-20\%.}

%% file: intro.tex
\vspace{-5pt}
\section{Introduction}

Binary similarity analysis determines if two given binary executables originate from the same source program. 
It has a wide range of applications such as automatic software patching~\cite{repair,shin2015recognizing,bao2014byteweight,bug-search-sem-sig,shari2021automated,zhengzi2017spain}, software plagiarism detection~\cite{clone-detection,binai,chae2013software,luo2017semantics,sajnani2016sourcerercc}, and malware detection~\cite{pmp,bilge2012disclosure, kharraz2015cutting,fratantonio2016triggerscope, kapravelos2014hulk,chan2016bingo, vulseeker}. For example, assume a critical security vulnerability has been reported and fixed in a library. It is of prominent importance to apply the patch to other deployed projects that included the library.
However, the library may be compiled with different settings in different projects. Binary similarity analysis allows identifying all the variants.
Given a pool of candidate binaries, which are usually functions in executable forms, a similarity analysis tool reports all the binaries in the pool equivalent to a {\em queried binary}. 
The problem is challenging as aggressive code transformations such as loop unrolling and function inlining in compiler optimizations may substantially change a program and produce largely different executables~\cite{bintuner}.

Given its importance, there is 
a large body of existing work. Earlier work (e.g., ~\cite{static-features,bindiff}) focuses on extracting static code features such as control-flow graphs and function call graphs. They are highly effective in detecting binaries that have small variations. 
Many proposed to use dynamic information instead~\cite{bin-match,blex,IMF,statistical-binsim} because it better discloses program semantics.
For example, {\em in-memory-fuzzing} (IMF)~\cite{IMF} uses fuzzing to generate many inputs and collects runtime information when executing the program on these inputs. It then uses the collected information to compute binary similarities. When the fuzzer can achieve good coverage, IMF is able to deliver high-quality results. However, achieving good coverage is difficult for complex programs (see our example in Section~\ref{sec:moti}).
Recently, Machine Learning and Deep Learning techniques are used to address the binary similarity problem~\cite{trex,binai,safe,How-Solve,kim2022improving,yang2022codee,diemph}. These techniques work by training models on a large pool of binaries that have positive and negative samples. The former includes binaries compiled from the same source and the latter includes those that are functionally different. The models are hence supposed to learn (implicit) features that can be used to cluster functionally equivalent programs. However, as shown in Sections~\ref{sec:limitation} and \ref{sec:eval:compare-to-baseline}, %
these models may learn features that are not robust, and in many cases, not semantics oriented, leading to sub-optimal results.

Inspired by the existing works that leverage dynamic information~\cite{IMF,bin-match,blex,osprey}, we consider the semantics of a binary to be a distribution of its inputs and their corresponding {\em externally observable values} during executions. Observable values are those encountered in I/O operations, and global/heap memory accesses. Compared to other runtime values such as those in registers, observable values are persistent across automatic code transformations as compilers hardly optimize these behavior~\cite{IMF,bin-match}. 
However, since we need to compare arbitrary binaries, ideally, we would have to collect sufficient samples in the input space of all these binaries. Making such samples universally comparable is highly challenging.
In Section~\ref{sec:limitation}, we show that a naive sampling strategy that executes all subject binaries on the same set of seed inputs can hardly work as different binaries take inputs of different formats.
For example, a valid input for a program $A$ is very likely an invalid input for programs $B$ and $C$. As such, it can only trigger similar error handling logics in $B$ and $C$, making them not distinguishable.

In this paper, we propose a sampling technique that can effectively approximate  semantics distributions by selecting and interpreting a small set of equivalent paths across different versions of a program. It is powered by a novel probabilistic execution engine. It runs 
candidate binaries on a fixed set of random seed values. Although many of these seed values lead to input errors, it systematically unfolds the program behavior starting from the execution paths of these seed values, called the {\em seed paths}. Specifically, it flips a bounded number of predicates along the seed paths. For instance, flipping a failing input check forces the binary to execute its normal functionality. 
\revise{While predicate flipping is not new~\cite{pmp,xforce, predicateswitch_zhang_2006}, 
our technique features a {\em probabilistic sampling algorithm}. Specifically, we cannot afford exhaustively exploring the entire neighborhood (of the seed paths) even with a small bound (of flipped predicates). %
Hence, we leverage a key observation that 
the predicates with the largest and the smallest {\em dynamic selectivity} tend to be stable before and after
automatic transformations, while other predicates vary a lot (by the transformations). Dynamic selectivity is a metric computed for a predicate instance that measures the distance to the decision boundary.
For example, assume a predicate {\tt x>y} yields true, {\tt x-y} denotes its dynamic selectivity.
Our theoretical analysis in Section~\ref{sec:formal-path} 
discloses that 
since automatic transformations cannot invent new predicates, but rather remove, duplicate, and reposition them, the likelihood that code transformations change the ranking of predicates with the smallest/largest selectivity is much smaller than that for other predicates.
Hence, we 
sample paths by flipping predicates that have close to the largest and the smallest selectivity, following the {\em Beta-distribution}~\cite{conti-dist} that has a U shape, biasing towards the two ends.}
Therefore, if two binaries are equivalent, our algorithm can sample a set of corresponding paths in the binaries by flipping their corresponding predicates such that the observable values along these paths disclose the equivalence. 

Our contributions are summarized as follows. 

\begin{itemize}
    \item We propose a novel probabilistic execution model that can effectively sample the semantics distribution of a binary and make the distributions from all binaries comparable.
    \revise{\item We develop a path sampling algorithm that is resilient to code transformation and capable of sampling equivalent paths when two binaries are equivalent. We also conduct a theoretical analysis to disclose its essence.
    \item We propose a probabilistic memory model that can tolerate invalid memory accesses due to predicate flipping while respecting the critical property of having equivalent behavior when the binary programs are equivalent.}
    \item We develop a prototype \toolname. We conduct experiments on two commonly used datasets including 35k functions from 30 binary projects and compare \toolname with five baselines~\cite{blex,IMF,trex,safe,How-Solve}. The results show that \toolname can achieve more than 90\% precision on average whereas the baselines can achieve 76\%. \toolname is also much more robust when the true positives (i.e., binaries equivalent to the queried binary)
    are mixed with various numbers of true negatives (i.e., binaries different from the queried binary) in the candidate pool, which closely mimics real-world application scenarios. Consequently, \toolname can correctly find 7 out of 8 1-day CVEs from binaries in the wild, whereas the baselines can only find 2. We upload \toolname at~\cite{PEM} for review. 
\end{itemize}

%% file: background.tex
\vspace{-5pt}
\section{Motivation and Overview}
\subsection{Motivating Example}
\label{sec:moti}

Our motivating example is adapted from the main function of \texttt{cat} in {\tt Coreutils}. 
The simplified source code is shown in Fig.~\ref{fig:src:cat}. %
Lines~\ref{line:cat:cli0} to~\ref{line:cat:cli1} parse the command line options. Lines~\ref{line:cat:print0} to~\ref{line:cat:print1} iteratively read the file names from the command line and emit the file contents to the output buffer. The function delegates the main operations to two functions. When some conditions at line~\ref{line:cat:condi} are satisfied, a simpler method {\it simple\_cat()} is called. Otherwise, it calls a more complex function that formats the output according to the full panoply of command line options. For example, at line~\ref{line:cat:print-inv}, if the 
global flag \texttt{print\_invisible} is set, the function prints out the ASCII values of invisible characters.

Compiler optimizations may substantially transform a program.
In Fig.~\ref{fig:cfg:cat0} and Fig.~\ref{fig:cfg:cat3}, we show the control flow graphs (CFGs) for our motivating example generated by two respective compilation settings,  -O0 meaning no optimization and 
-O3 meaning having all commonly used optimizations applied.
The {\tt switch} statement at line~\ref{line:cat:switch}  is compiled to hierarchical {\tt if-then-else} structures with -O0, as shown in the orange circle in Fig.~\ref{fig:cfg:cat0}. In contrast, it is compiled to an {\it indirect jump} with -O3, as shown in the orange circle in Fig.~\ref{fig:cfg:cat3}.
The predicate at line~\ref{line:cat:condi} corresponds to the blue circle in Fig.~\ref{fig:cfg:cat0}. We can see two branches, each consisting of only one basic block. Two delegated functions are called in the two basic blocks, respectively.
However, the two functions are inlined in the optimized version, resulting in branches with much more blocks, e.g., 50 blocks in the branch of the complex function, as shown in the blue circle in Fig.~\ref{fig:cfg:cat3}.
To better illustrate the challenges, we introduce another function adapted from the main function of {\tt touch} in {\tt Coreutils}, as shown in Fig.~\ref{fig:src:touch}. The function {\tt touch} modifies the meta information of files.
Lines~\ref{line:touch:optbegin} to~\ref{line:touch:optend} parse the command line options and the {for}-loop at line~\ref{line:touch:touchbegin} iteratively performs the touch operation. We can see from Fig.~\ref{fig:cfg:touch0} that the syntactic structures of {\tt touch} and {\tt cat} are more similar than those between {\tt cat} with and without optimizations. %
The observation can be quantified by the statistics of these CFGs shown in the caption.

\input{fig_tex/fig_motivation.tex}
\input{fig_tex/fig_cfgs.tex}

\vspace{-5pt}
\subsection{Limitations of Existing Techniques}
\label{sec:limitation}
\noindent \textbf{Fuzzing-Based Techniques.} There are techniques that leverage fuzzing to explore the dynamic behavior of programs and use them in similarity analysis. For example, {\it in-memory fuzzing} (IMF)~\cite{IMF} iteratively mutates function inputs and collects traces.
Since the parameter specifications for functions in stripped binaries are not available, it is challenging to generate inputs that can achieve good coverage.
In our example, IMF can hardly generate legal command line options for the function {\tt main\_cat}. Thus most collected behavior is from the error processing code at line~\ref{line:cat:error}. Moreover, they tend to collect similar (error processing) behavior from {\tt main\_touch}.
As such, the downstream similarity analysis likely draws the wrong conclusion about their equivalence. 
Our experiments in Section~\ref{sec:eval:compare-to-baseline} show that IMF can achieve a precision of 76\% on complex cases, whereas ours can achieve 96\%.%

\revise{
\noindent \textbf{Forced-Execution-Based Techniques.} To extract more behavior from binary code, there are methods that use coverage as guidance to execute every instruction in a brute-force fashion. A representative work BLEX~\cite{blex} executes a function from the entry point. Then it iteratively selects the first unexecuted instruction to start the next round of execution until every instruction is covered. We call techniques of such nature forced-execution-based as they largely ignore path feasibility. 
\rebuttalshort{
There are two essential challenges for these techniques. First, they tend to use code coverage within a function as the guidance for forced execution, which has the inherent difficulty in dealing with function inlining~\cite{bintuner}. Another challenge is to provide appropriate execution contexts when execution starts at arbitrary (unexecuted) locations. For example, suppose that in the first few rounds, BLEX executes the  true branch at line~\ref{line:cat:simple} of Fig.~\ref{fig:moti_ex}. When it tries to cover the false branch at line~\ref{line:cat:complex}, it uses a fresh execution context, discarding the variables computed at line~\ref{line:cat:define}. 
}
According to our experiments in Section~\ref{sec:eval:compare-to-baseline}, these techniques can achieve a precision
of 69\%, whereas our technique can achieve 96\%. %
}

\begin{figure*}
    \centering
    \begin{subfigure}[t]{.28\textwidth}
        \includegraphics[width=\textwidth]{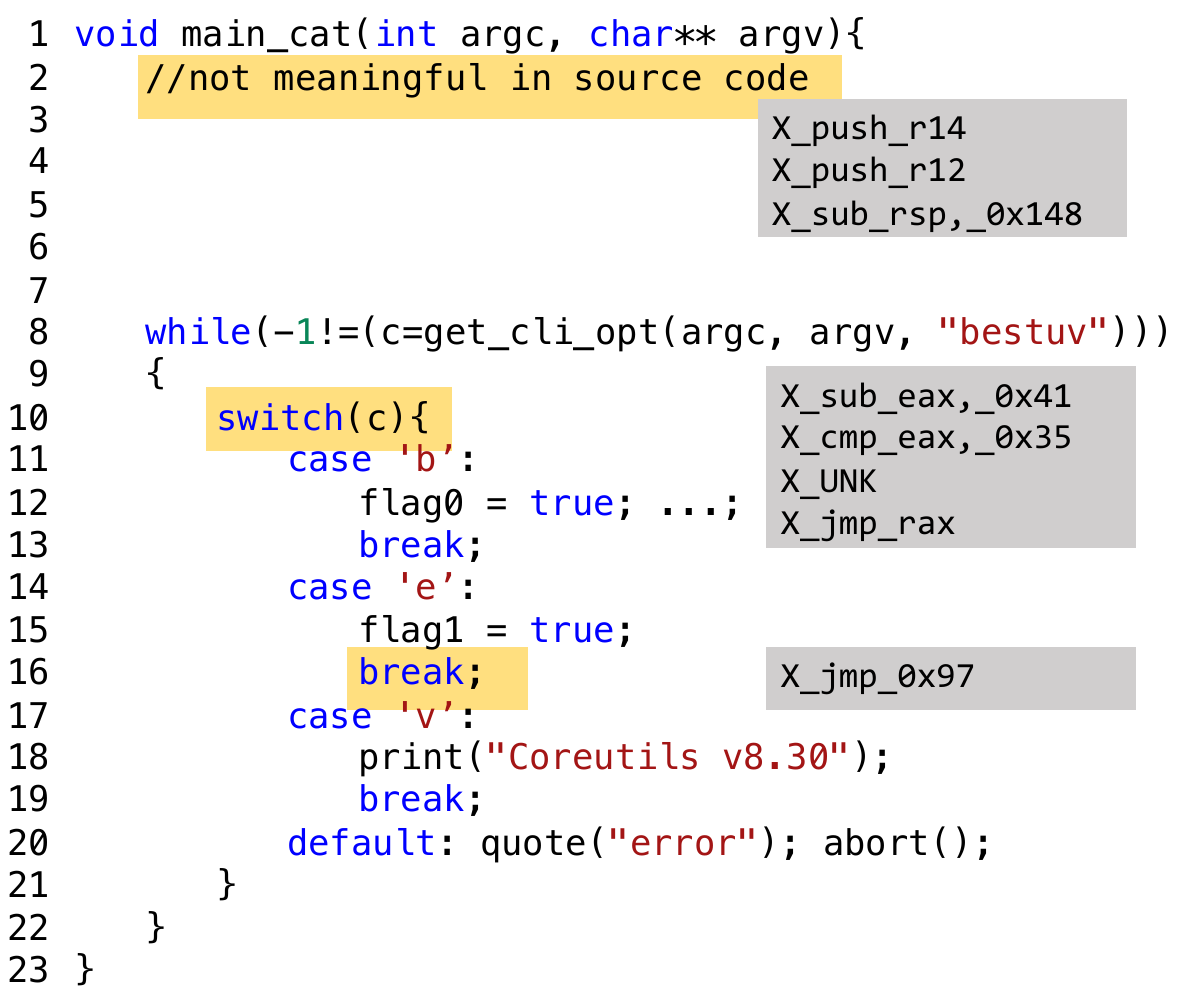}
        \vspace{-30pt}
        \caption{cat@O3}
        \label{fig:att:cat3}
    \end{subfigure}
    \rulesep
    \begin{subfigure}[t]{.28\textwidth}
        \includegraphics[width=\textwidth]{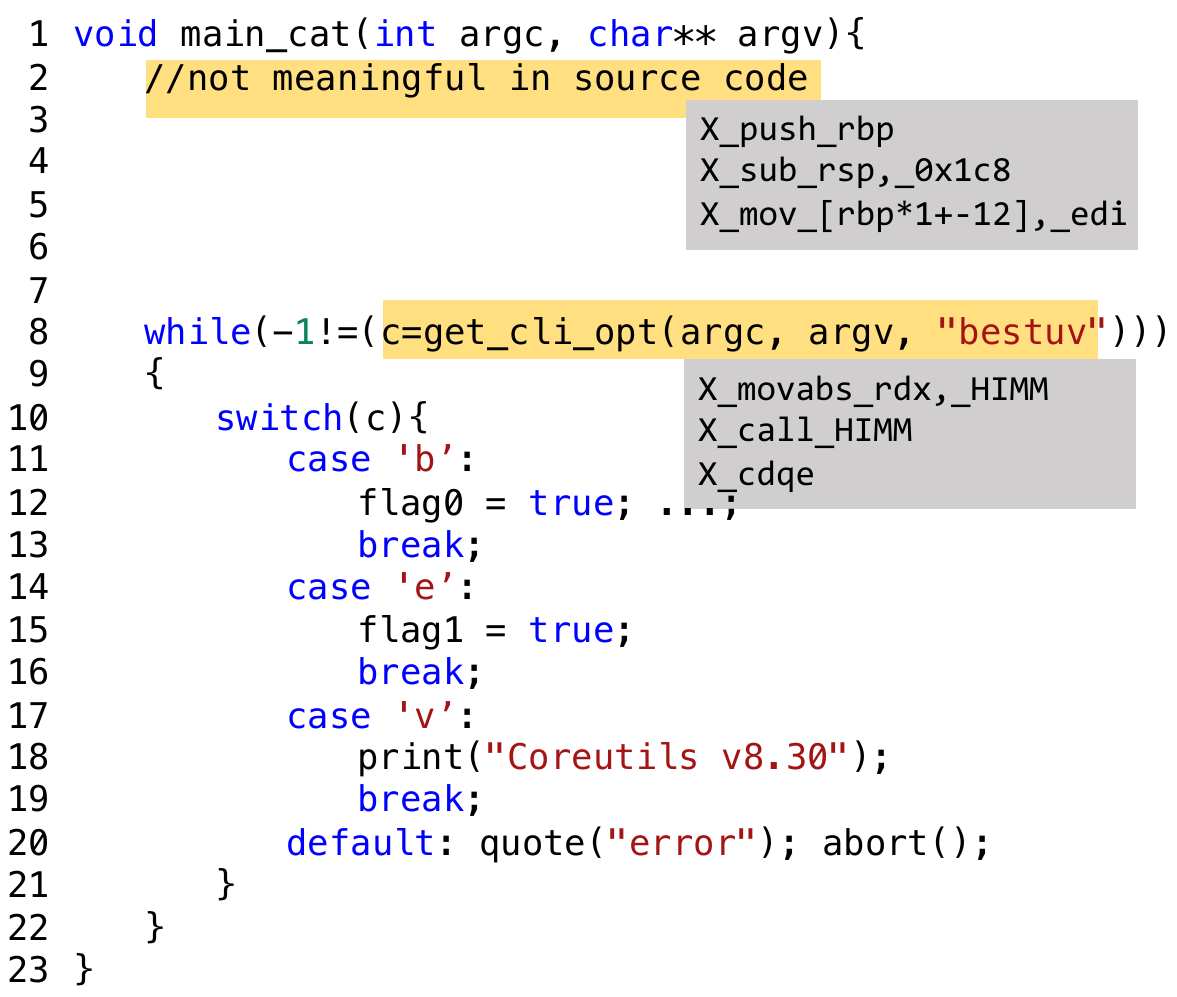}
        \vspace{-30pt}
        \caption{cat@O0}
        \label{fig:att:cat0}
    \end{subfigure}
    \rulesep
    \begin{subfigure}[t]{.28\textwidth}
        \includegraphics[width=\textwidth]{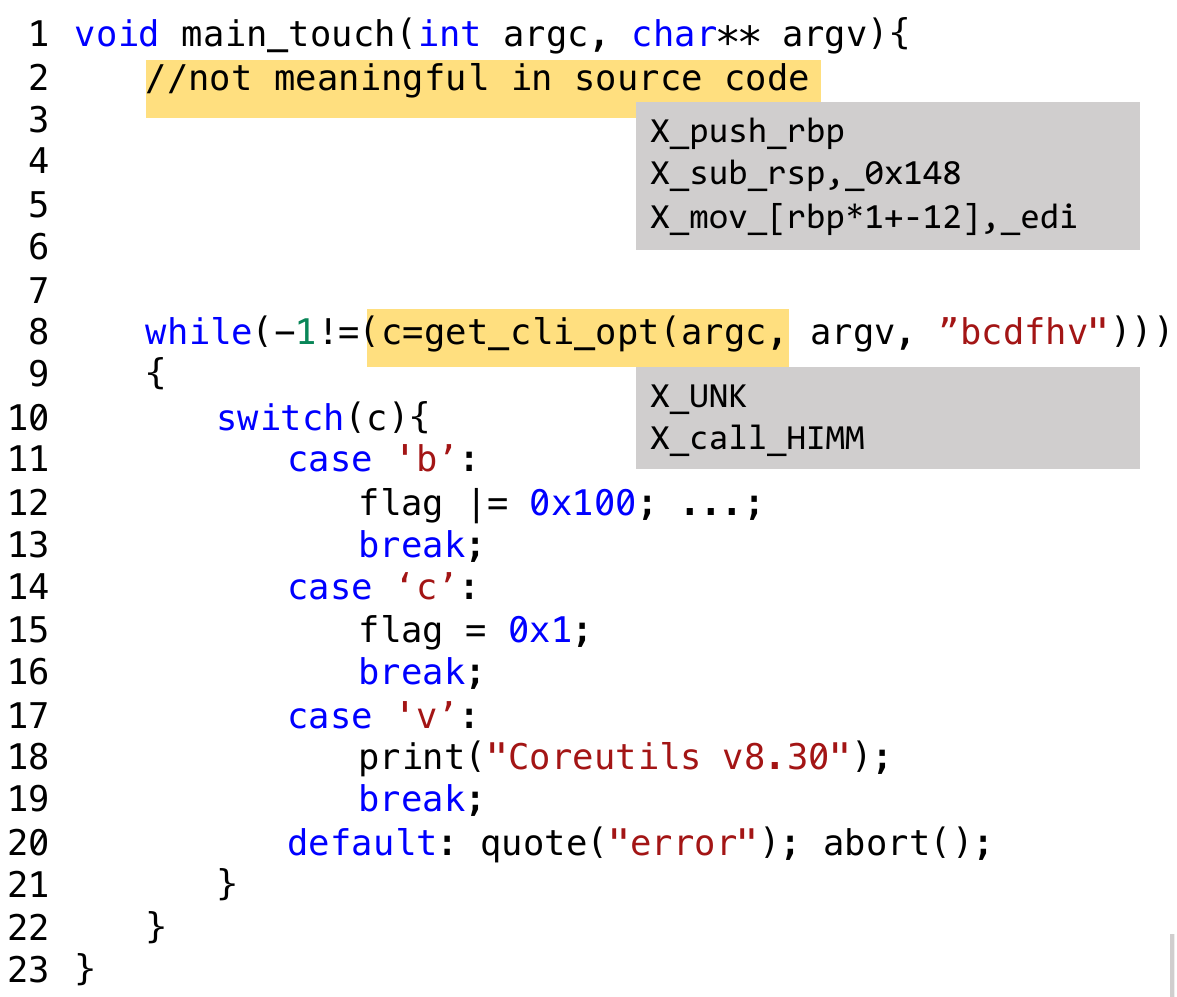}
        \vspace{-30pt}
        \caption{touch@O0}
        \label{fig:att:touch0}
    \end{subfigure}
    \caption{Our example (Fig.~\ref{fig:moti_ex}) in  SAFE. The statements highlighted in yellow have large attention (and hence are important). The gray boxes to the right (of the yellow statements) denote the corresponding tokens. Special token {\tt HIMM} denotes a constant or a constant control flow target. %
    }
    \label{fig:vis_att}
    \vspace{-15pt}
\end{figure*}

\noindent \textbf{Learning-Based Techniques.} Emerging techniques~\cite{trex,safe,binai,gmn} leverage Machine Learning models. Some models~\cite{binai,gmn} extract static features from CFGs. However, these static features are not robust in the presence of optimizations.
Another line of work uses language models~\cite{trex,safe}. Their hypothesis is that these models could learn instruction semantics and hence function semantics.
To limit the vocabulary (i.e., the set of words/tokens supported), binaries are often normalized before they can be fed to models. For example, immediate values (i.e., constants in instructions) and constant call targets are replaced with a special token {\tt HIMM} in SAFE~\cite{safe}, e.g., the token {\tt x\_call\_HIMM} around line 8 in Fig.~\ref{fig:vis_att} (b) and (c) that corresponds to the function invocation {\tt get\_cli\_opt}. 
While this makes training convergence feasible, a lot of semantics are lost. 

\rebuttalshort{
These models may not learn to classify based on instructions essential to function semantics.
For example, SAFE leverages an NLP technique called {\it attention}~\cite{attention}. Conceptually, the attention mechanism determines which instructions are important to the output. 
We highlight the statements and their tokens with the largest attention values in Fig.~\ref{fig:vis_att}.
}
In these three functions, the first few tokens (in gray) with large attention values are in the function prologues. The corresponding instructions  (e.g., {\tt push}) perform the same functionality,  saving register values to memory and allocating space for local variables. 
In Fig.~\ref{fig:att:cat3}, the model also pays attention to tokens/instructions related to the switch-case statement. As discussed before, however, static structures are not reliable due to optimizations. In contrast, in Fig.~\ref{fig:att:cat0}, the model instead emphasizes the normalized function invocation at line 8, which is not distinguishable from the invocation at line 8 in (c) with a large attention value as well.
From the parts that the model pays attention to,
it is easy to explain why SAFE concludes {\tt cat@O0} is more similar to {\tt touch@O0}, instead of {\tt cat@O3}. We visualize the weights of full attention layers in 
Fig.~\ref{fig:apdx:att} 
of the appendix.

\vspace{-5pt}
\subsection{Our Technique}
\label{sec:moti:tech}
We aim to leverage program semantics in similarity analysis. We define the semantics of a binary program $P$ as follows.
\vspace{-4pt}
\begin{definition}
The semantics of a binary program $P$ is represented by a distribution 
$(x, OV({P(x)}))\sim \mathcal{D}$, with $x\in \mathcal{X}$ an input to $P$ and $OV(P(x))$ the set of externally observable values when executing $P$ on $x$. Observable values are those observed in I/O operations, global, and heap memory accesses.  
\end{definition}
\vspace{-4pt}
Intuitively, the joint distribution of inputs and observable values when executing $P$ on the inputs denotes $P$'s semantics. 
Observable values are hardly altered by code  transformations.

\noindent
{\bf A Naive Sampling Method.}
One may not need to collect a large number of samples to model the aforementioned distribution because if two programs are equivalent, executing them on equivalent inputs produces equivalent observable values. 
Therefore, a naive method is to provide the same set of inputs to all programs such that those that are equivalent must have identical observable value distributions.  
However, such a simple method is ineffective because of the following reasons.
First, even equivalent programs might have different input specifications\;(e.g., different numbers of parameters and different orders of parameters),\;making automatically feeding equivalent inputs to them difficult.
Furthermore, different programs have different input domains. When the provided inputs are out-of-range (and hence invalid), the corresponding observable value distributions cannot be used to cluster programs. In our example, the valid domain of $c$ at line~\ref{line:cat:switch} of {\tt main\_cat} is a set of characters {\it\{b,e,s,t,u,v\}} whereas the domain of $c$ at line~\ref{line:touch:switch} of {\tt main\_touch} is {\it\{b,c,d,f,h,v\}}. Without input specifications, which are hard to acquire for binary functions, the naive sampling method may provide a random input, say, $c\!=\!173$. As a result, the executions of both functions fall into the exception handling paths and the observable values are not distinguishable. 

\noindent
{\bf Our Method.}
Instead of solving the input specification problem, which is very hard for binary programs, we propose a technique agnostic to such specifications. 
Specifically, we propose a novel probabilistic execution model that serves as an effective sampling method to approximate the distribution $\mathcal{D}$ denoting program semantics.
Given a program $P$, we acquire its semantic representation as follows. We execute $P$ on a set $\mathcal{X}$ of pre-determined (random) inputs, which is an invariant for all programs we want to represent.
To address the challenge of input specification differences, we assign the same value $x\in \mathcal{X}$ to each input variable  (for all programs). That is, we feed the same value to all input parameters, making their order irrelevant.
We repeat this for all values in $\mathcal{X}$.
As an example, 
for the programs in Fig.~\ref{fig:moti_ex},
we set {\tt argc} and {\tt **argv} (all elements in the buffer) in both {\tt main\_cat} and  {\tt main\_touch}, as well as {\tt *inbuf}, {\tt insize}, and {\tt *outbuf} in function {\tt complex\_cat}, to 173, acquiring three executions. Then we set them to 97, \textls[-15]{acquiring another three executions}, and so on. %

These random values may not be valid inputs and hence the corresponding executions may not disclose meaningful semantics. 
We hence further sample {\em k-edge-off} \ behavior.
\vspace{-3pt}
\revise{
\begin{definition}
Given a program $P$ and an input $x$, let $p$ be the program path taken with input $x$, we say a path $p'$ is $k$-edge-off (from $p$) if$\ $ $k$ predicates along the execution need to be flipped to other branch outcomes in order to acquire $p'$.
\end{definition}
\vspace{-3pt}
For instance, suppose that when executing {\tt main\_cat} with $c=173$, the path $p$ is \ref{line:cat:cli0}-\ref{line:cat:switch}-\ref{line:cat:error}. If the branch at line~\ref{line:cat:cli0} is flipped to line~\ref{line:cat:define},
assuming that the following execution path is
\ref{line:cat:condi}-\ref{line:cat:malloc}-\ref{line:cat:complex}-\ref{line:cat:while},
\ref{line:cat:cli0}-\ref{line:cat:define}-\ref{line:cat:condi}-\ref{line:cat:malloc}-\ref{line:cat:complex}-\ref{line:cat:while} is $1$-edge-off from $p$.
$K$-edge-off behavior (of an input $x$) is essentially the observable values encountered in all $k$-edge-off paths (of $x$).
Observe that for {\tt main\_cat} and {\tt main\_touch}, although the 0-edge-off behaviors (i.e., the original executions) are not distinguishable, the 1-edge-off behaviors are quite different, e.g., the behavior of {\tt main\_cat} includes those from the delegated function at line~\ref{line:cat:complex}.
}
However, there is a practical challenge: covering all $k$-edge-off behavior even when $k=2$ may be infeasible for complex programs since the number of $k$-edge-off paths grows exponentially with $k$. Moreover, controlling the sampling process exclusively by $k$ induces substantial noise due to code optimizations/transformations. Specifically, optimizations substantially change program structures, adding/removing predicates. 
The $k$-edge-off behaviors are hence quite different. 
An example can be found in 
Section~\ref{supp:moti-example}
of the appendix.
To suppress the noise introduced by optimizations, we leverage the observation that optimizations rarely
change the (selectivity) ranking of predicates with  the maximum and minimum {\em dynamic selectivity}.
\vspace{-3pt}
\revise{
\rebuttal{
\begin{definition}
Dynamic selectivity for a predicate instance $x \otimes y$ is $\big|[\![y]\!]-[\![x]\!]\big|$,
where $[\![y]\!], [\![x]\!]$ are the runtime values of variable $x$ and $y$, and $\otimes \in \{>, \ge, \ne, ==, <, \le \}$. %
\end{definition}
}
\vspace{-3pt}
For instance, suppose that in an execution, the value of {\tt inbuf[i]} at line~\ref{line:cat:print-inv} in Fig.~\ref{fig:moti_ex} is 173. It is then compared with 0x20. 
The dynamic selectivity of the predicate instance is
hence  141 (i.e., $| 173-\text{0x20}|$).
Essentially, the dynamic selectivity reflects how likely a branch predict evaluates to true~\cite{preach}.
Although automatic code transformations may change dynamic selectivity, 
the predicate instances with the largest/smallest dynamic selectivity tend to stay as the largest/smallest ones after transformations. %
We formally explain the observation in Section~\ref{sec:formal-path} 
and empirically validate it in Section~\ref{sec:eval:abl}. Therefore, we select predicate instances to flip following a {\em Beta-distribution}~\cite{conti-dist} with $\alpha=\beta=0.03$.
The distribution has the largest probabilities for predicates with the minimum and maximum selectivity and small probabilities in the middle (like a U shape).  
Intuitively, if two programs are equivalent/similar, their predicates with the largest and the smallest selectivity tend to be the same. By flipping these predicates in the two versions, we are exploring their equivalent new behavior.

In our example, for both the optimized and the unoptimized version of {\tt main\_cat}, the algorithm first flips the predicate at line~\ref{line:cat:cli0} with a high probability since the  {\tt -1!=c} has the largest selectivity on path \ref{line:cat:cli0}-\ref{line:cat:switch}-\ref{line:cat:error}. Then we achieve  the 1-edge-off path \ref{line:cat:cli0}-\ref{line:cat:define}-\ref{line:cat:condi}-\ref{line:cat:malloc}-\ref{line:cat:complex}-\ref{line:cat:while} 
as discussed above. Along the new path, the algorithm flips the predicate with the largest selectivity at line~\ref{line:cat:print-inv} for further 2-edge-off exploration in both versions, exposing similar behavior.}

To realize the probabilistic execution model, we develop a binary interpreter that can feed the binary with specially crafted inputs and sample observable values (Section~\ref{sec:interp}). It also features a probabilistic memory model that can tolerate invalid memory accesses while ensuring equivalent observable values for equivalent programs %
(Section~\ref{sec:PM}). 
Compared to traditional forced-execution-based techniques, \toolname naturally handles the function inlining problem as our sampling is not delimited by function boundaries and our execution contexts are largely realistic. 
Compared to fuzzing based techniques, ours does not rely on solving the hard problem of generating valid inputs.
Compared to Machine Learning based techniques, our technique focuses on dynamic behavior of programs, which are more accurate reflections of program semantics~\cite{kargen2017towards}.

%% file: fig_tex/fig_motivation.tex
\begin{figure}[t!]
\begin{subfigure}[b]{.48\textwidth}
\begin{minipage}{\textwidth}
\begin{lstlisting}[deletekeywords={input}, style=customc, firstnumber=1, xleftmargin=0.5cm, basicstyle=\scriptsize,
    escapeinside={(|}{|)}]
void main_cat(int argc, char** argv){
    while(-1!=(c=get_cli_opt(argc, argv, "bestuv"))){ (|\label{line:cat:cli0}|)
        switch(c){ (|\label{line:cat:switch}|)
            case 'b': flag0 = true; ...; format = true; break; (|\label{line:cat:f0}|)
            case 'e': flag1 = true; break;
            case 'v': print("Coreutils v8.30"); break;
            ...
            default: quote("error"); abort(); (|\label{line:cat:error}|)
        }
    }(|\label{line:cat:cli1}|)
    ...//define: (|\texttt{pageSize, inbuf}|) and (|\texttt{insize}|)  (|\label{line:cat:define}|)
    do{(|\label{line:cat:print0}|)
        if((flag0||flag1) && format){ (|\label{line:cat:condi}|)
            ret = simple_cat(inbuf, insize); (|\label{line:cat:simple}|)
        }else{ ...
            outbuf = xmalloc(outbuf, pageSize) (|\label{line:cat:malloc}|)
            ret = complex_cat(inbuf, insize, outbuf); (|\label{line:cat:complex}|)
        }
    }while(...);(|\label{line:cat:while}|)(|\label{line:cat:print1}|)
}
int complex_cat(char* inbuf, int insize, char* outbuf){
    if(print_invisible && inbuf[i]<0x20) (|\label{line:cat:print-inv}|)
        outbuf[...] = to_ascii(inbuf[i])
}
\end{lstlisting}
\end{minipage}
\vspace{-15pt}
\caption{Coreutils: cat}
\vspace{-3pt}
\label{fig:src:cat}
\end{subfigure}

\begin{subfigure}[b]{.48\textwidth}
\begin{minipage}{\textwidth}
\begin{lstlisting}[deletekeywords={input}, style=customc, firstnumber=1, xleftmargin=0.5cm, basicstyle=\scriptsize,
    escapeinside={(|}{|)}]
void main_touch(int argc, char** argv){
    while(-1!=(c=get_cli_opt(argc, argv, "bcdfhv"))){ (|\label{line:touch:optbegin}|)
        switch(c){ (|\label{line:touch:switch}|)
            case 'b': flag |= 0x100; break;
            case 'c': flag = 0x1; break;
            case 'v': print("Coreutils v8.30"); break;
            ...
            default: quote("error"); abort(); (|\label{line:touch:error}|)
        }
    }(|\label{line:touch:optend}|)
    for(int i = begin; i < argc; i++) (|\label{line:touch:touchbegin}|)
        touch_function(argv[i]); (|\label{line:touch:touchfunc}|)
    
}
\end{lstlisting}
\end{minipage}
\vspace{-20pt}
\caption{Coreutils: touch}
\label{fig:src:touch}
\end{subfigure}
  \caption{
  Motivating Example
  }
  \label{fig:moti_ex}
\end{figure}

%% file: fig_tex/fig_cfgs.tex
\begin{figure*}
\centering
\begin{subfigure}[b]{.26\textwidth}
    \includegraphics[width=\textwidth]{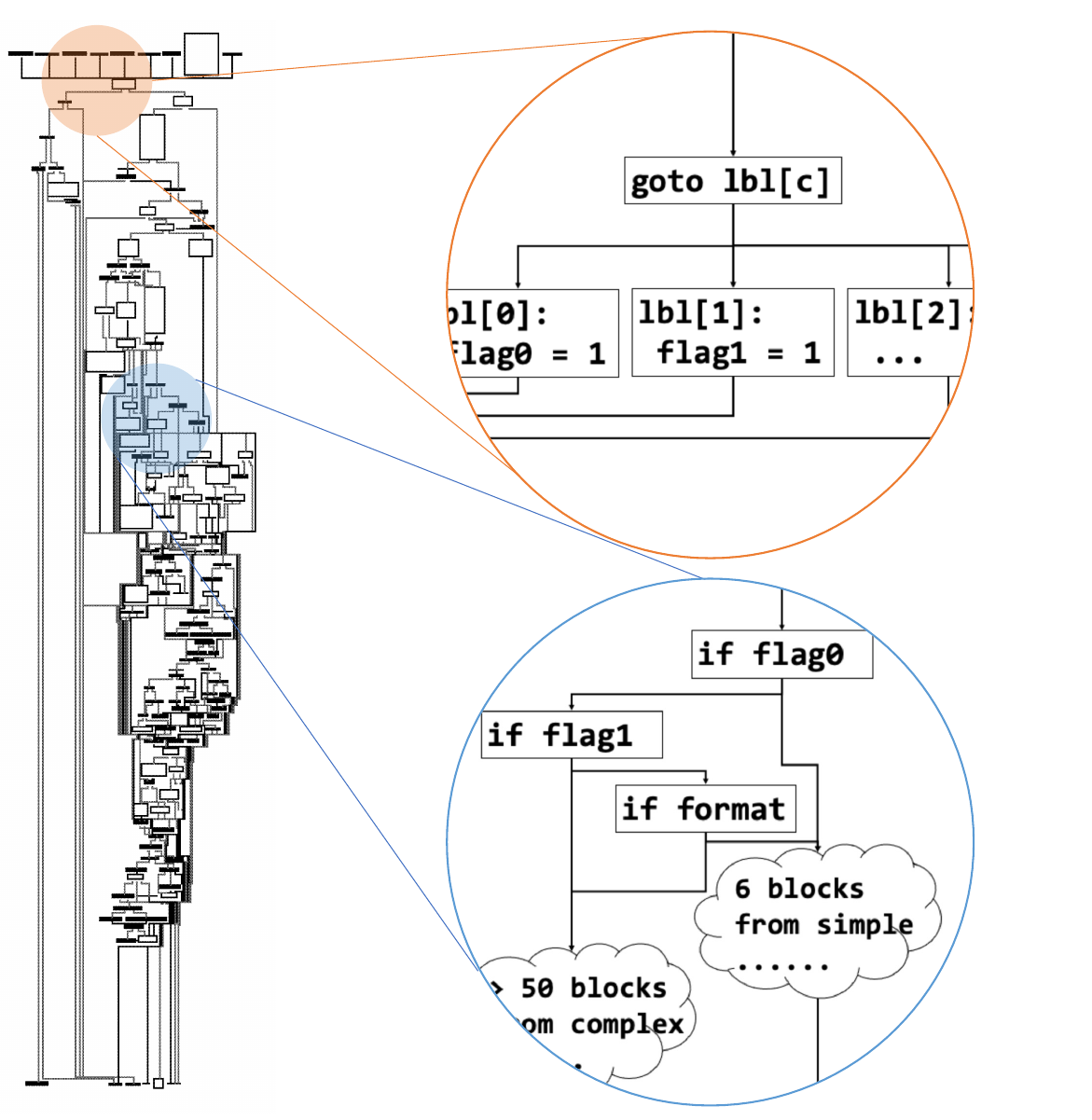}
    \vspace{-20pt}
    \caption{Cat@O3 (144 blocks and 218 edges)}
    \label{fig:cfg:cat3}
\end{subfigure}
\hspace{.01\textwidth}
\begin{subfigure}[b]{.26\textwidth}
    \includegraphics[width=\textwidth]{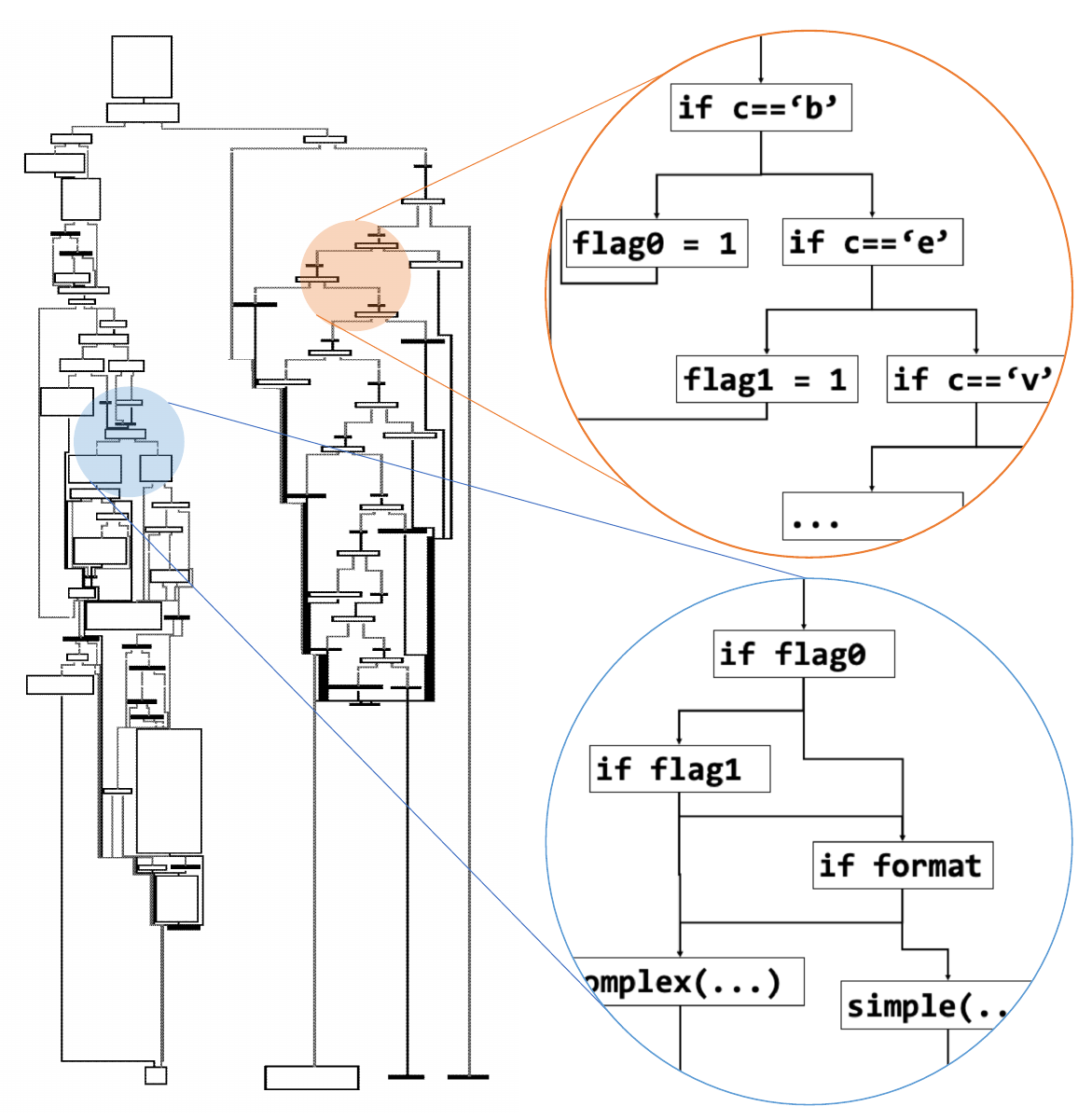}
    \vspace{-20pt}
    \caption{Cat@O0 (83 blocks and 144 edges)}
    \label{fig:cfg:cat0}
\end{subfigure}
\hspace{.01\textwidth}
\begin{subfigure}[b]{.26\textwidth}
    \includegraphics[width=\textwidth]{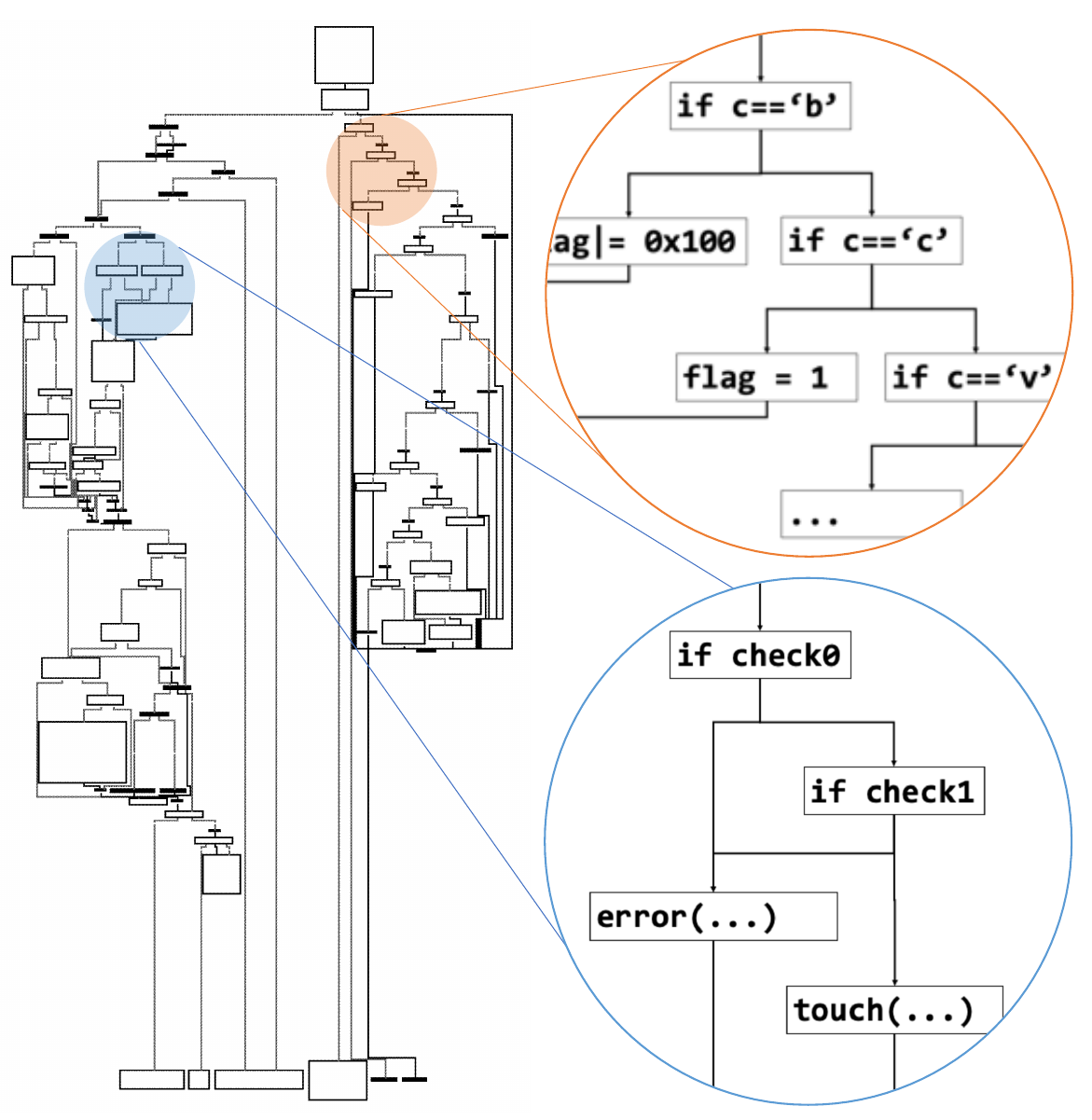}
    \vspace{-20pt}
    \caption{\textls[-20]{Touch@O0\,(89\,blocks~and~120\,edges)}}
    \label{fig:cfg:touch0}
\end{subfigure}

    \caption{Control-Flow-Graphs for Motivation Examples
    }
    \label{fig:cfgs}
    \vspace{-17pt}
\end{figure*}

%% file: method.tex
\newcommand{\langkw}[1]{\textbf{\color{blue}#1}}

\begin{figure}
    \centering
    \includegraphics[width=0.98\linewidth]{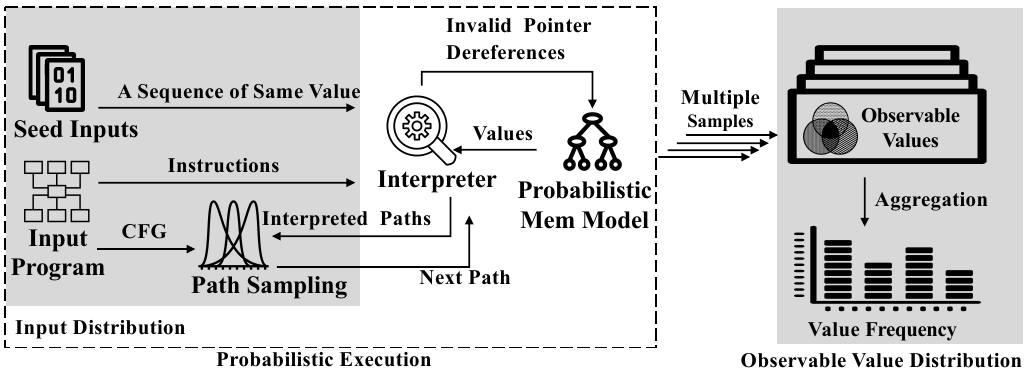}
    \caption{Workflow of \toolname
    }
    \label{fig:wf}
    \vspace{-3pt}
\end{figure}

\vspace{-5pt}
\section{Design}
\label{sec:design}
\vspace{-3pt}
\subsection{Overall Workflow}
\vspace{-1pt}
The workflow of \toolname is shown in Fig.~\ref{fig:wf}. The input is in the grey box on the left side. It consists of a set of seed inputs, each being an infinite sequence of the same value, the binary executable, and a path sampling strategy that can predict the next path to interpret based on the set of interpreted paths.
The interpreter interprets the 
subject binary on a seed input, supplying the same value to any input variable encountered during interpretation, to eliminate any semantic differences caused by parameter order differences.
The interpretation also strictly follows the path indicated by the path sampling component. 
When invalid pointer dereferences are encountered, which can be easily detected, the interpreter interacts with the probabilistic memory model to emulate the access outcomes. The emulation ensures that the same sequence of (observable) values are returned for equivalent paths.
After sampling, on the right side, the observable value distributions are summarized for later similarity analysis, which simply compares two multi-sets.

The remainder of this section is organized as follows. 
We first model binary instructions using a simplified language. Then we present the semantic rules. After that, we discuss the path sampling method and the probabilistic memory model.

\vspace{-5pt}
\subsection{Language}
\vspace{-1pt}
The syntax of our language is in Fig.~\ref{fig:definition}. A program $P$ consists of a sequence of instructions. There are three categories of instructions. First, there are instructions that move values among registers: $r_1=r_2$ moves the value in $r_2$ to $r_1$; $r=v$ moves a literal value $v$;
$r_1 = r_2 \diamond r_3$ moves the result of $r_2 \diamond r_3$ to $r_1$.
The second category is load and store instructions. The load instruction
$r_1 = [r_2]$ treats the value in $r_2$ as a memory address and loads the value in the specified memory location to $r_1$. 
Store is similar.
There are also instructions that change the control flow. Instruction $\langkw{jmp}\ a$ jumps to the instruction at $a$; $\langkw{jcc}\ r\ a$ performs the jump operation only when the value in $r$ is non-zero; $\langkw{jr}\ r$ is an indirect jump that uses the value in $r$ as the target address.
Instruction $\langkw{done}$ means the interpretation is finished. 
Although our language does not model functions for simplicity, our implementation supports the full x86 instruction set, including function invocations and returns. %

\vspace{-3pt}
\subsection{Interpretation}
\label{sec:interp}
\vspace{-1pt}
The state domains of the interpreter are illustrated in the upper box of Fig.~\ref{fig:abs}. 
The register state $Regs$ is a mapping from a register to a value.
While in our presentation values are simply non-negative integers, our implementation distinguishes bytes, words, and strings. 
The memory store $M$ is a mapping from an address to a value.
We use an instruction counter $ic$ to identify each interpreted instruction along the execution path. 
$OV$ denotes the observable value statistics. 
It is a mapping from a value to the number of its observations, that is, how many times the value appears in the current interpretation.
In the lower box, we define a number of auxiliary data/structures that are immutable during interpretation and a number of helper functions used in the semantic rules.
In particular, we use $\bot$ to denote an undefined  value; $p$ to denote the path to interpret, determined by the path sampling component (for a given seed value $s$).
It is a mapping from instruction count to an instruction address. For example, a 2-edge-off path for a seed value 994 can be $\{1000\rightarrow \text{0x804578},\ 2000\rightarrow \text{0x80a41f}\}$. It means that the predicate instance with the instruction count 1000 ought to take the branch starting at 0x804578 when executing the binary with the seed input 994, and the instance with count 2000 should take the branch at 0x80a41f.  
The helper function {\bf decode}($a$) disassembles the instructions in a basic block starting at $a$.
The function {\bf valid}($a$) determines if an address is valid. Note that since we enforce branch outcomes and use crafted inputs, the execution states may be corrupted. This function helps detect such corrupted states and seeks help from the probabilistic memory model.
The function {\bf invalidLd}($a$) loads a value from an invalid address.

\input{fig_tex/fig_syntax.tex}

\begin{figure}[t]
    \begin{mdframed}
      {\footnotesize \tt
      $Regs\!\in$\texttt{RegState}$\Coloneqq$$\texttt{Register}\!\rightarrow\! \texttt{Val}$ \\
      $M\!\! \in$\;\texttt{Memory}\;$\Coloneqq$\;$\texttt{Addr}\!\! \rightarrow\!\! \texttt{Val}$\;\;\;\;\;\;\;\;\;\;\;\;\;\;\;\;\;\;\;\;\;
      $ic\!\in$\texttt{InstrCnt}$\Coloneqq$$\mathbb{Z}+$\;
      $OV\!\!\in$\texttt{ObservableValDist}$\Coloneqq$$\texttt{Val}\!\! \rightarrow\!\!\mathbb{Z}+$
      }
    \end{mdframed} \vspace{-13pt}
    \begin{mdframed}
    \setlength\parindent{8pt}
    {\footnotesize
    $\bot$ $\Coloneqq$ \texttt{Undefined Value} \;\;\;\;\;\;\;\;\;\;
     $p \in$ \texttt{Path} $\Coloneqq$ $\texttt{InstrCnt} \rightarrow \mathbb{Z}+$\\
      $s \in$ \texttt{SeedValue} $\Coloneqq$ $\mathbb{Z}+$\\
    \texttt{\textbf{decode}($a$)}: returns instructions in a basic block starting from $a$.\\
    \texttt{\textbf{valid}($a$)}: if an address $a$ is valid (pointing to allocated memory). \\
    \texttt{\textbf{invalidLd}($a$)}: load a value from an invalid address  $a$. %

  }
  \vspace{-3pt}
    \end{mdframed}
    \caption{State Domains in Interpretation (top) and Auxiliary Data and Functions (bottom)
    }
    \label{fig:abs}
\end{figure}

Part of the semantic rules are in Fig.~\ref{fig:sem_rule}. 
As shown at the top of Fig.~\ref{fig:sem_rule}, the state configuration is a tuple of five entries. 
A rule is read as follows: if the preconditions at the top are satisfied, the state transition at the bottom takes place.
For example, Rule {\it JccGT} says that if %
there is a branch $a'$ specified in $p$ for the current instruction count $ic$, the conditional jump is interpreted and the continuation is $I'$ decoded from $a'$. 
\input{fig_tex/fig_rules.tex}

Intuitively, given a seed value $s$, the interpreter initializes all registers and parameters with the same value $s$, and starts interpretation from the beginning (Rule {\it Start}). The interpretation largely follows concrete execution semantics except the following. First, when it encounters a conditional jump which is indicated by the path descriptor $p$ to take a specific branch, it takes the specified branch (Rule {\it JccGT}). Otherwise, it follows the normal semantics (Rules {\it JccT} and {\it JccF}). Second, when it encounters 
a load, if the address is valid but the memory location has not been defined, it fills it with $s$ (Rule {\it LdUd}); if the address is invalid, it fetches a value from the probabilistic memory model (Rule {\it LdIv}); otherwise it loads a value from the memory as usual (Rule {\it LdV}). Here
$Regs'=\update{Regs}{r_1}{v}$ means that the register state is updated by associating $r_1$ to $v$, yielding a new state $Regs'$. 
Store instructions are interpreted similarly. %
We track all dynamic memory allocations for access validity checks. Details are elided as this is standard.

We also have a set of logging rules that describe how \toolname{} records the statistics of observable values.
We record the frequencies of memory addresses accessed, values loaded/stored, control transfer targets, and predicate outcomes.
Due to space limitations, details are presented in 
Section~~\ref{supp:log-rule}
of the appendix.

\noindent
{\bf Loops and Recursion.}
Since our goal is to disclose semantic similarity and not to infer semantics faithful to any executions induced by real inputs, 
following common practice, we unroll each loop and recursive call 20 times.

\vspace{-5pt}
\subsection{Path Sampling}
\label{sec:path}
\vspace{-3pt}

\input{method-path-sampling}

\vspace{-3pt}
\subsection{Probabilistic Memory Model}
\vspace{-3pt}
\label{sec:PM}

The goal of the probabilistic memory model (PMM) 
is to handle loads and stores with invalid addresses induced by predicate flipping and the use of (out-of-bound) seed values.
A key observation is that  the specific values written-to/read-from the PMM do not matter as long as they can expose functional equivalence.~We~define the following two properties for a valid~PMM.
\begin{definition}
\vspace{-5pt}
We say a PMM is {\em equivalence preserving} if the sequence of (invalid) addresses accessed, and the values written-to/read-from the PMM must be equal, for two equivalent paths in two functionally equivalent programs. 
\end{definition}
\vspace{-5pt}
This property ensures \toolname can place equivalent programs into the same class.
\vspace{-5pt}
\begin{definition}
We say a PMM is {\em difference revealing} if the sequence of (invalid) addresses accessed, and the values written-to/read-from the PMM must be different for two different paths (pertaining invalid memory accesses) in two respective programs, which may or may not be equivalent. \end{definition}
\vspace{-5pt}
This is to ensure different programs are not mistakenly placed in the same class. 
For example, a naive PMM always returns a constant value for any invalid reads and ignores any invalid writes. It is equivalence preserving but not difference revealing. \looseness=-1

\textls[-10]{
Our PMM is designed as follows. Before each interpretation run, it initializes a {\em probabilistic memory} ($PM$), which is a mapping $\texttt{Addr}\rightarrow \texttt{Val}$ of size $\gamma$ such that:
$\forall a\in [0,\gamma],\ PM[a]=random()$.
An invalid memory read from the normal memory $M$ with address $a$ is forwarded to the $PM$ through the {\bf invalidLd}($a$) function, which returns $PM[a\mod \gamma]$.
Similarly, an invalid memory write to the normal memory $M$ with address $a$ and value $v$ is achieved by 
setting $PM[a\mod \gamma]=v$.}

It can be easily inferred that our PMM satisfies the equivalence preserving property by induction (on the length of program paths). Intuitively, the first invalid accesses in two equivalent paths must have the same invalid address. As such, our PMM must return the same random value. This same random value may be used to compute other identical (invalid) addresses in the two paths such that the following invalid loads/stores are equivalent.
It also probabilistically satisfies the difference revealing property. Specifically, different paths manifest themselves by some different invalid addresses, and our PMM likely returns different (random) values for these different addresses, rendering the following memory behaviors (with invalid addresses) different. The chance that different paths may exhibit the same behavior depends on $\gamma$. Due to the complexity of modeling memory behavior in real-world program paths, we did not derive a theoretical probabilistic bound for our PMM. However, empirically we find that $\gamma=64k$ enables very good results (with our loop unrolling bound 20). An example can be found in Section~\ref{supp:example-pmm} of the appendix.

%% file: fig_tex/fig_syntax.tex
\begin{figure}[t]
\vspace{-5pt}
    \begin{mdframed}
      {\footnotesize \tt
        \addtolength{\tabcolsep}{-5pt}
  \begin{grammar}
    <\texttt{Program}> $P$ $\Coloneqq$ $I$
\ \ 
    $\langle\texttt{Register}\rangle$ $R$ $\Coloneqq$ $\{r_0, r_1, \dots, r_{31}\}$  \vspace{-5pt}
    
    <\texttt{Val}> $v$ $\Coloneqq$ $\{0, 1, 2, ...\}$
    $\;\;\;\;\;\;\;\;\;\;\;\;\;$
    $\langle\texttt{Addr}\rangle$ $a$ $\Coloneqq$ $\{0, 1, 2, ...\}$
    \vspace{-5pt}
    
    <\texttt{Comparison}> $\otimes$ $\Coloneqq$ $\{==, \ge, \le, \dots\}$ \;\;\;\;
    %
    $\langle\texttt{BinOp}\rangle$ $\diamond$ $\Coloneqq$ $\{+, -, *, \dots\}$
    \vspace{-5pt}
    
    <\texttt{Instruction}> $I$ $\Coloneqq$ $r_1=r_2$|$r = v$|$r_1 = r_2 \otimes r_3$|$r_1 = r_2 \diamond r_3$ 
    \alt $r_1 = [r_2]$ | $[r_1] = r_2$  | $\langkw{jmp}\ a$ | $\langkw{jcc}\ r\ a$ | $\langkw{jr}\ r$
    | $\langkw{done}$
    | $I_1;I_2$
    
  \end{grammar}
      }
    \end{mdframed}
    \caption{Syntax of Our Language}
    \label{fig:definition}
    \vspace{-10pt}
\end{figure}

%% file: fig_tex/fig_rules.tex
\newcommand{\update}[3]{\ensuremath{#1[#2\rightsquigarrow#3]}\xspace}
\newcommand{\progstate}[5]{\ensuremath{\langle #1,#2,#3,#4,#5\rangle}\xspace}
\begin{figure}[t]
\begin{mdframed}
\begin{minipage}{\textwidth}
{\footnotesize
\vspace{-3pt}
\boxed{\textbf{State Configuration:}\  \progstate{Regs}{I}{ic}{M}{OV}} \\
\vspace{-3pt}
\begin{mathpar}
\inferrule{\texttt{EntryPoint} = en\\ \forall r, Regs[r]=s
}
{\progstate{\emptyset}{\emptyset}{0}{\emptyset}{\emptyset}
\rightarrow
\progstate{Regs}{\textbf{decode}(en)}{1}{\emptyset}{\emptyset}
}
{Start}\\
\vspace{-3pt}
\inferrule{
p[ic]=a' \\ I'=\textbf{decode}(a')
}
{\progstate{Regs}{\langkw{jcc}\ r\ a;I}{ic}{M}{OV}
\rightarrow
\progstate{Regs}{I'}{ic+1}{M}{OV}
}
{JccGT}\\
\vspace{-3pt}%
\inferrule{Regs[r]\ne 0\\ p[ic]=\bot \\ I'=\textbf{decode}(a)}
{\progstate{Regs}{\langkw{jcc}\ r\ a;I}{ic}{M}{OV}
\rightarrow
\progstate{Regs}{I'}{ic+1}{M}{OV}
}
{JccT}\\
\vspace{-3pt}
\inferrule{Regs[r]=0 \\ p[ic]=\bot}
{\progstate{Regs}{\langkw{jcc}\ r\ a;I}{ic}{M}{OV}
\rightarrow
\progstate{Regs}{I}{ic+1}{M}{OV}
}
{JccF}\\
\vspace{-3pt}
\inferrule{Regs[r_2]=a\\  v=M[a]\\ Regs'=\update{Regs}{r_1}{v}}
{\progstate{Regs}{r_1 = [r_2];I}{ic}{M}{OV}
\rightarrow
\progstate{Regs'}{I}{ic+1}{M}{OV}
}
{LdV}\\
\vspace{-3pt}
\inferrule{Regs[r_2]=a\\ \textbf{valid}(a) \\ M[a]=\bot\\ Regs'=\update{Regs}{r_1}{s}}
{\progstate{Regs}{r_1 = [r_2];I}{ic}{M}{OV}
\rightarrow
\progstate{Regs'}{I}{ic+1}{M}{OV}
}
{LdUd}\\
\vspace{-3pt}
\inferrule{Regs[r_2]=a\\ \neg\textbf{valid}(a) \\
v=\textbf{invalidLd}(a)\\
Regs'=\update{Regs}{r_1}{v}}
{\progstate{Regs}{r_1 = [r_2];I}{ic}{M}{OV}
\rightarrow
\progstate{Regs'}{I}{ic+1}{M}{OV}
}
{LdIv}
%
%
\end{mathpar}
} 
\end{minipage}
\label{fig:sem:rule_exec}
\end{mdframed}    
\caption{Interpretation Rules 
}
\label{fig:sem_rule}
\vspace{-3pt}
\end{figure}

%% file: method-path-sampling.tex
We present the path sampling method in Algorithm~\ref{alg:path-sampling}.
It consists of two functions. 
Function {\it interpret} at line~\ref{alg:line:interpret_fun} interprets the input program and flips the predicates that are indicated by {\it path}, a mapping from instruction count to an address (Fig.~\ref{fig:abs}). 
Specifically, during interpretation, the algorithm flips a predicate instance to an address indicated in {\it path} if the corresponding instruction count is met.
The function returns a list of encountered predicate instances. 

Function {\it sample} 
iteratively selects a predicate instance to flip (from all the interpretation results in previous steps). 
Variable {\it candidates} denote a set of candidate predicates for flipping and {\it budget} the number of interpretations allowed.
To begin with, \toolname first interprets a faithful path without altering any branch outcome. It then adds predicates in this faithful path to the candidates list (line~\ref{alg:line:addcandi}). \looseness=-1

\begin{algorithm}[h]

\revise{
\caption{Probabilistic Path Sampling}\label{alg:path-sampling}
\Func{interpret(program, path)}{\label{alg:line:interpret_fun}
\tcp{\textnormal{Interprets the input program; flips predicates indicated by} {\it path}}
\vspace{-1pt}
\tcp{\textnormal{Returns predicate instances in the form {\it(path, instruction count, predicate,  selectivity, outcome)}}}
\vspace{-1pt}
\Return{[$(path, ic_0, pr_0, sel_0, out_0), ... $]}
}
\vspace{-3pt}
\Func{sample(program)}{\label{alg:line:sample}
\vspace{-1pt}
    candidates = []\tcp{\textnormal{Candidate branches to flip}} \label{alg:line:candi}
    \vspace{-1pt}
    budget = 400 \tcp{\textnormal{Number of sample rounds}} \label{alg:line:budget}
    \vspace{-1pt}
    faithful = interpret(program, $\emptyset$)\\
    \vspace{-1pt}
    candidates.add(faithful)\\ \label{alg:line:addcandi}
    \vspace{-1pt}
    \While{budget $\ge$ 0}{\label{alg:line:loop}
    \vspace{-1pt}
        budget = budget - 1\\
        \vspace{-1pt}
        (path,ic,pr,sel,out) = select(candidates)\\ \label{alg:line:toflip}
        \vspace{-1pt}
        nextPath = path $\cup$ \{ic $\rightarrow$  getBranch(pr,$\neg$out)\}\\ \label{alg:line:nextpd}
        \vspace{-1pt}
        results = interpret(program, nextPath)\\\label{alg:line:interpret}
        \vspace{-1pt}
        candidates.add(results)\\ \label{alg:line:addnew}
        \vspace{-1pt}
    }
        \vspace{-1pt}
}
}
\vspace{-4pt}
\end{algorithm}
As shown in the loop at line~\ref{alg:line:loop}, 
\toolname iteratively selects a predicate to flip~(line~\ref{alg:line:toflip}), composes a new path with the outcome of the selected predicate flipped at line~\ref{alg:line:nextpd} (function {\it getBranch()} acquires the target address for the true/false branch outcome of a predicate {\it pr}), interprets the program according to the new path~(line~\ref{alg:line:interpret}), and updates the list of candidates~(line~\ref{alg:line:addnew}). 
\rebuttal{
Note that at line~\ref{alg:line:toflip}, to select the predicate instance to flip,
\toolname sorts all the candidate predicates by their dynamic selectivity. Then a real number $i\in[0,1]$ is sampled following the {\it probability density function} (PDF) of a Beta-distribution~\cite{conti-dist}. \toolname selects a predicate that is at the $i$-percentile of the sorted candidates list, i.e., $selected\_pred = sorted\_candidates[i\times (len(sorted\_candidates)-1)]$.
Details can be found in Section~\ref{supp:impl} of the appendix.
}

\vspace{-5pt}
\subsection{Formal Analysis of Path Sampling}
\label{sec:formal-path}
\vspace{-1pt}
The effectiveness of our path sampling algorithm piggybacks on the following theorem.
\vspace{-5pt}
\begin{theorem}
\textls[-12]{
Assume two functionally equivalent programs $P$ and $P'$. If we interpret them along two equivalent paths and collect the predicate instances during interpretation, the predicate instances with the largest (smallest) dynamic selectivity in both programs have a larger probability to match, compared to those with non-extreme selectivity.\looseness=-1}
\end{theorem}
\vspace{-5pt}

While optimizations (e.g., {\em constraint elimination}~\cite{llvm}) may modify predicates to simplify control flow,  predicates with the smallest and largest dynamic selectivity are most resilient to optimizations, namely, their selectivity ranking hardly changes before and after optimizations.
Modifications to predicates introduced by optimizations fall into two categories: {\em predicate elimination} and {\em insertion}.
A predicate relocation can be considered as first removing the predicate and then adding it to another location.
Specifically, compiler may eliminate a predicate if its outcome is implied by the path condition reaching the predicate. For example, it may eliminate a predicate $x>10$ if the path condition includes $x>20$. 
On the other hand, compiler may introduce new predicates to provide control flow shortcuts. Take Fig.~\ref{fig:opt} as an example. 
Compiler inserts a new predicate, $x<10$, in Fig.~\ref{fig:opt:after-opt}~(shown in red). The modification simplifies the control flow when $x$ is less than $10$.
Note that, in these cases, the dynamic selectivity of an inserted predicate will be close to the dynamic selectivity of an existing one because these inserted predicates are derived from constraints in existing predicates.

\begin{figure}[h]
\centering
\begin{subfigure}[b]{.43\linewidth}
\begin{minipage}{\textwidth}
\begin{lstlisting}[deletekeywords={input}, style=customc, firstnumber=1, xleftmargin=0.5cm, basicstyle=\footnotesize,
    escapeinside={(|}{|)}]
if x == 10: ...
else if x == 15: ...
else if x == 20: ...
...
else: abort()
\end{lstlisting}
\end{minipage}
\vspace{-9pt}
\caption{Before Optimization}
\label{fig:opt:before-opt}
\end{subfigure}
\begin{subfigure}[b]{.43\linewidth}
\begin{minipage}{\textwidth}
\begin{lstlisting}[deletekeywords={input}, style=customc, firstnumber=1, xleftmargin=0.5cm, basicstyle=\footnotesize,
    escapeinside={(|}{|)}]
(|\color{red}\texttt{if x < 10: abort() }|)(|\label{line:opt:insert}|)
if x == 10: ...
else if x == 15: ...
else if x == 20: ...
...
else: abort()
\end{lstlisting}
\end{minipage}
\vspace{-9pt}
\caption{After Optimization}
\label{fig:opt:after-opt}
\end{subfigure}
\caption{Example of optimization that provides control flow shortcut by inserting predicates. The compiler inserts a predicate {\tt x<10} at line~\ref{line:opt:insert} in Fig.~\ref{fig:opt:after-opt}. When {\tt x<10}, the execution directly goes to {\tt abort()} without comparing with other values.
  }
  \label{fig:opt}
\end{figure}
The intuition of our theorem is hence that the rankings of  predicates with the smallest/largest selectivity do not depend on whether other predicates are modified.
In contrast, the predicates ranked in the middle by their selectivity are more likely to have their rankings changed when predicates are removed or added by optimization.

\begin{figure}[t]
    \centering
    \includegraphics[width=.45\linewidth]{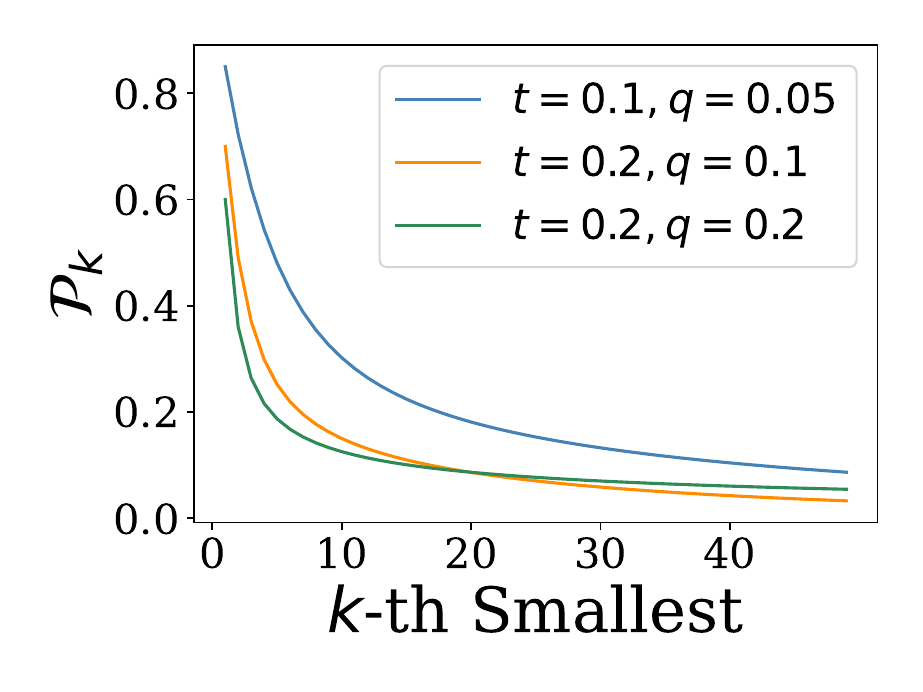}
    \vspace{-5pt}
    \caption{$\mathcal{P}_k$ w.r.t. $k$; The $x$-axis denotes the ranking of predicates by dynamic selectivity; the $y$-axis denotes the probability that the predicate with the $k$-th smallest dynamic selectivity after optimization has the same ranking. Each line shows results for one set of $t$ and $q$.\looseness=-1}
    \label{fig:prob-stable}    
    \vspace{-1pt}
\end{figure}

\smallskip
\noindent
{\em Proof Sketch.}
\textls[-6]{
We formalize the intuition by first reasoning about the predicates having close to the smallest dynamic selectivity. Reasoning for the largest ones is symmetric.
Suppose that for each predicate, the compiler has a probability $t$
to eliminate it and a probability $q$ for having a predicate inserted that ranks right before it.
In either case, we say the predicate is modified.
The probability that a predicate is not modified is noted as $r=1-t-q$. We further denote as $\mathcal{P}_k$ the probability that the $k$-th smallest predicate is still the $k$-th smallest one after optimization. It is calculated by the following formula:}
\vspace{-5pt}
\begin{equation}
\vspace{-5pt}
 \label{eqt:p_stable}
  \mathcal{P}_k = r\times \sum_{i=0}^{\lfloor\frac{k-1}{2}\rfloor}{{k-1 \choose 2i}{2i \choose i}r^{k-1-2i}t^iq^i}
\end{equation}
 Intuitively, the ranking of the $k$-th smallest predicate is not changed by optimizations if (a)~this predicate is not modified and (b)~the number of predicates with a smaller dynamic selectivity does not change.
 In the above formula, $r$ represents condition~(a) and the second term represents condition~(b). Specifically,
 (b) is satisfied only when the numbers of removed and inserted predicates  that rank before $k$ are equal.  Here,  
 ${k-1 \choose 2i} r^{k-1-2i}$ means an even number ($2i$) of the $k-1$  predicates with a smaller ranking are modified, and  ${2i \choose i}t^iq^i$ means half of the modifications are removals and the other half are insertions.
 We visualize the distribution of $\mathcal{P}_k$ in Fig.~\ref{fig:prob-stable} with three sets of configurations of $t$ and $q$. We can see that in all setups, $\mathcal{P}_k$ monotonically decreases when $k$ increases. $\Box$
 
\smallskip
\noindent
We also conduct an empirical study to validate our theoretical analysis. The results are visualized in Section~\ref{sec:eval:abl}.
The results show \toolname has an 80-90\% chance of making correct selections and exploring equivalent paths by deterministically selecting the predicates with largest/smallest dynamic selectivity.

\smallskip
\noindent
{\bf Advantages of Probabilistic Path Sampling Over Deterministic Selection.}
Note that the probability of predicates with the smallest/largest selectivity having their rankings changed by optimization is not 0, although it is smaller than others. 
To tolerate such certainty, we employ a probabilistic approach, meaning that we follow a Beta-distribution instead of deterministically selecting the predicates with extreme selectivities for flipping.
We further conduct a formal analysis to justify why the probabilistic sampling algorithm is better  than the deterministic algorithm. 
Intuitively, by following a Beta distribution, PEM spends some budget on predicates that do not have the largest or smallest selectivity, but  selectivities close to the largest and smallest. These ``additional'' selections increase the probability that PEM selects the correct path (i.e., the equivalent path) at each step. Taking more correct steps at earlier selections increases the chance that PEM chooses a correct step at later selections because the candidate predicates of later selections come from previously explored paths.
The formal proof is shown in Section~\ref{supp:theoretical} of the appendix.

\smallskip
\noindent
{\bf Effect of Path Infeasibility.}
Our algorithm may select infeasible paths. Two possible concerns are (1) whether observable values along infeasible paths in two similar binaries can correctly disclose their semantic similarity; and (2) whether observable values along infeasible paths  
in two dissimilar binaries may undesirably match, leading to the wrong conclusion of their similarity.

\textls[-8]{
For the first concern, 
we show that \toolname likely selects corresponding paths when two binaries are similar, regardless of the feasibility of selected paths. That is, although the paths may be infeasible, the sequences of observable values along them are equivalent. 
We show a proof sketch in Section~\ref{appdx:proof-path-1} and show empirical support in Section~\ref{appdx:path-emp} of the appendix.\looseness=-1 }

For the second concern, the probability that two equivalent paths are selected by \toolname in two dissimilar binaries is very small. In those cases, although the initial seed paths may be undesirably similar  (e.g., the error handling paths), the following flipped (infeasible) paths quickly become substantially different.
The formal proof is in Section~\ref{appdx:proof-path-2} and the empirical study is in Section~\ref{appdx:path-emp} of the appendix. %

%% file: eval.tex
\vspace{-4pt}
\section{Evaluation}
\label{sec:eval}
\vspace{-3pt}

We implement \toolname on QEMU~\cite{qemu}. Details are in Section~\ref{supp:impl} of the appendix.~\looseness=-1
We evaluate \toolname via the following research questions:\\
\textbf{RQ1:} How does \toolname perform compared to the baselines?\\
\textbf{RQ2:} How useful is \toolname in real-world applications?\\
\textbf{RQ3:} Is \toolname generalizable?\\
\textbf{RQ4:} How does each component affect the performance?

\vspace{-4pt}
\subsection{Setup}
\vspace{-3pt}
We conduct the experiments on a server with a 24-core Intel(R) Xeon(R) 4214R CPU at 2.40GHz, 188G memory, and Ubuntu 18.04.

\noindent
{\bf Datasets.} 
We use two datasets.
 {\em Dataset-\MakeUppercase{\romannumeral 1}}: To compare with IMF and BLEX, which only use Coreutils~\cite{coreutils} as their dataset, we construct a dataset from Coreutils-8.32. We compile the dataset using GCC-9.4 and Clang-12, with 3 optimization levels (i.e., -O0, -O2, and -O3). 
{\em Dataset-\MakeUppercase{\romannumeral 2}} includes 9 real-world projects commonly-used in binary similarity analysis projects~\cite{trex,How-Solve,binkit}. They are Coreutils, Curl, Diffutils, Findutils, OpenSSL, GMP, SQLite,  ImageMagick, and Zlib. The binaries are obtained from~\cite{trex}. 
In total, we have 30 programs with 35k functions, compiled with 3 different options. 
Details can be found in Table~\ref{tab:dataset-stats} of the appendix. 

\noindent
{\bf Baseline Tools.} 
We compare with 6 baselines.
For execution-based methods (Baseline-I), we use IMF~\cite{IMF} and BLEX~\cite{blex}, which are SOTAs as far as we know.
For Deep Learning methods (Baseline-II), we use SAFE~\cite{safe} and Trex~\cite{trex}. We use their pre-trained models or train using their released implementation with the default hyper-parameters. 
Also, we compare with the best two models (i.e., GNN and GMN) in How-Solve~\cite{How-Solve} that conducts a measurement study on  Machine Learning methods. 

\noindent{\bf Metrics.} 
Following the same experiment setup 
in IMF and BLEX, for a function compiled with a higher level optimization option (e.g., -O3), we query the most similar function in all the functions (in the same binary) compiled with a lower level optimization option.
As such, there is only one matched function. 
We hence use Precision at Position 1 (PR@1) as the metric. Given a function, PR@1 measures whether the matched function scores the highest out of the pool of candidate functions. %
Many data-driven methods~\cite{binai,trex,safe,How-Solve}  use {\em Area Under Curve} (AUC) of the {\em Receiver Operating Characteristic} (ROC) curve.
Existing literature~\cite{DoandDont} points out that a good AUC score does not necessarily imply good performance 
on an imbalanced dataset (e.g., 
class 1 having 1 sample and class 2 having 100). Therefore we choose PR@1 as our metric,
which aligns better with the real-world (imbalanced) use scenario of binary similarity.

\input{tab_tex/tab_eva_baseline.tex}

\vspace{-3pt}
\subsection{RQ1: Comparison to Baselines}
\vspace{-3pt}
\label{sec:eval:compare-to-baseline}
\noindent{\bf Comparison to Baseline-\MakeUppercase{\romannumeral 1.}}
We compare \toolname with Baseline-\MakeUppercase{\romannumeral 1} on Dataset-\MakeUppercase{\romannumeral 1}.
To conduct the evaluation, we first use \toolname to sample each function in these binaries and aggregate the distribution of observable values. 
Then, for each function in an optimized binary, we compute its similarity score against all functions of the same program compiled with a lower optimization level, and use the ones with the highest scores to compute PR@1. 
Besides PR@1, we also use PR@3 and @5 
for a more thorough comparison with IMF.
\textls[-7]{
The comparison results with IMF and BLEX are shown in Table~\ref{tab:baseline1}\footnote{We compare \toolname with the reported results in the IMF paper and contact the authors of IMF to ensure our setups are the same.}. The first two columns list the compilers and the optimization flags used to generate the reference and query binaries. Columns 3-5, 6-7, and 8-9 list PR@1, @3, and @5, respectively. 
Note that BLEX only reports PR@1 and does not have results for binaries compiled with Clang.
\toolname outperforms BLEX on PR@1 and outperforms IMF on all 3 metrics under all settings. Especially, for function pairs (Clang-O0, GCC-O3) and (GCC-O0, Clang-O3), which are the most challenging settings in our experiment, \toolname outperforms IMF by about 25\%. 
}

\noindent{\bf Comparison to Baseline-\MakeUppercase{\romannumeral 2}.} We compare \toolname with Baseline-\MakeUppercase{\romannumeral 2} on Dataset-\MakeUppercase{\romannumeral 2}.
Following the setup of How-Solve~\cite{How-Solve}, for each positive pair (of functions), namely, similar functions, 100 negative pairs (i.e., dissimilar functions) are introduced to build up the test set. The results are shown in Fig.~\ref{fig:how-help}.
The $x$ axis represents different programs, and the $y$ axis is PR@1. The results of \toolname, GNN, and GMN are shown in green, yellow, and red bars. The average PR@1 of each tool is marked by the dashed line with the related color. Note that GNN and GMN are the best two models out of all 10 ML-based methods in How-Solve~\cite{How-Solve} (including Trex and SAFE).
As Fig.~\ref{fig:how-help} illustrates, \toolname achieves scores from 0.84 to 1.00, which is around 20-40\% better than GNN and GMN.\looseness=-1

\noindent
{\em Comparison with Trex and SAFE.}
With the aforementioned composition of dataset, \toolname outperforms Trex by 40\% and outperforms SAFE by 25\% on average. Moreover,
the performance of Trex and SAFE is sensitive to dataset composition. 
Hence in this comparative experiment, we  analyze how different data compositions affect the performance of different tools.
Our results show that \toolname is 50\% more resilient than Trex and SAFE. Details can be found in Section~\ref{supp:ratio-perf} of of the appendix. 

\input{fig_tex/fig_eva_how_solve.tex}
\vspace{-3pt}
\subsection{RQ2: Real-World Case Study}
\vspace{-3pt}
\textls[-5]{
We demonstrate the practice use of \toolname via a case study of detecting 1-day vulnerabilities. 
Suppose that after a vulnerability is reported, a system maintainer 
wants to know if the vulnerable function occurs in a production system. %
She can use \toolname to search for the vulnerable function from a large number of binary functions and decide whether further actions should be taken (e.g., %
patch the system).
We collect 8 1-day Vulnerabilities (CVEs) and  use the optimized version of the problematic function to search for its counterpart in the unoptimized binary.
The results show that in 7 out of the 8 cases, our tool can find the ground truth function as the top one, while the other two ML-based methods each can only find 1 of them.
Even if we look into the top 30, both of them can only find 2 of these problematic functions.
Details can be found in Section~\ref{supp:cve} of the appendix.}
\vspace{-4pt}
\subsection{RQ3: Generalizability}
\vspace{-3pt}
We evaluate the generalizability of \toolname from three perspectives. First, we show that \toolname is efficient so that it can scale to large projects. Second, we illustrate that \toolname has good code coverage for most functions. That means it can explore enough semantic behavior even for complex functions. Last but not least, besides x86-64, we show that \toolname can support another architecture with reasonable human efforts, meaning that \toolname can be easily generalized to  analyzing binary programs from multiple architectures, without the need of substantial efforts in building lifting or reverse engineering tools to recover high-level semantics from binaries.

\revise{
\noindent
{\bf Efficiency.} 
\toolname analyzes more than 3 functions per second in most cases. Note that 
this is a one-time effort. After interpretation and generating  semantic representations,
\toolname searches these representations to find similar functions.
\toolname compares more than 2000 pairs per second in most cases. %
The comparison can be parallelized. 
With 4 processes, we are able to compare 1.7 million function pairs in 4 minutes (wall-clock). We visualize the results in Figure~\ref{fig:time-eff} of the appendix.
}
\input{fig_tex/fig_eva_coverage.tex}

 \toolname takes 13 minutes to cover more than 95\% code for all functions in Coreutils (with a single thread). In comparison, the forced-execution based method BLEX takes 1.2 hours.
 In our experiment, \toolname takes 26 minutes to process two Coreutils binaries compiled with different optimization levels, and it takes another 14 minutes to compare all 1.7 million function pairs between these two binaries, yielding a total time cost of 40 minutes. While IMF takes 32 minutes to complete the same task, \toolname achieves significantly better precision than IMF.
Machine learning models typically have an expensive training time. They have better performance in test time.

\noindent
{\bf Coverage.} The code coverage of \toolname on Dataset-\MakeUppercase{\romannumeral 2} is shown in Fig.~\ref{fig:Coverage}. The $x$ axis marks the projects and the $y$ axis shows the percentage of functions for which \toolname has achieved various levels of coverage, denoted by different colors. As we can see, 90\% of the functions in  -O0 and 85\% functions in  -O3 have a full or close-to-full coverage. Those functions with less than 40\% coverage have extremely complex control flow structures, with 
many inlined callees.
For example, the main function of {\it sort} in {Coreutils} has 496 basic blocks, resulting in  millions of potential paths. Note that even with such a huge path space, \toolname is still able to select similar paths and collect consistent values with a high probability.

\revise{
\noindent
{\bf Cross-arch Support.} We add AArch64~\cite{arm64} support to \toolname with only around 200 lines of C++ code and 0.5 person-day efforts. This is possible because our probabilistic execution model is general and does not rely on specialized features from the underlying architecture.
\toolname achieves a PR@1 of 86.8 for Coreutils (-O0 and -O3) on AArch64,
whereas its counterpart on x86-64 is 89.4. %
In addition, it achieves a PR@1 of 84.9 when we query with functions 
compiled on x86-64 in the pool of functions
compiled on AArch64. Details can be found in Table~\ref{tab:multi-arch} of the appendix.
}
\vspace{-4pt}
\subsection{RQ4: Ablation Study}
\label{sec:eval:abl}
\vspace{-3pt}
\revise{

\noindent
{\bf Probabilistic Path Sampling.} First, we empirically validate our hypothesis that branches with the largest and smallest selectivity are stable before and after code transformations. We collect equivalent interpretation traces from the main functions in Coreutils binaries compiled with different options. Then we 
analyze the matching traces %
and check if the predicates with the largest and the smallest selectivity in these cross-version traces match, leveraging the debug information. 
In total, we study 636 traces from 6 binaries with a total of 16k predicate instances. 
We observe that with a probability of more than 80\%, our hypothesis holds. The detailed results are shown in Fig.~\ref{fig:branch-relation}.
From the two ends of the lines, we can observe that in more than 80\% cases, the predicates with the smallest and the largest selectivity match. In contrast, those in the middle do not have such a property.
The median for the {\it max-3} selectivity is even close to 0\%. %
\textls[-4]{
For predicate instances with the smallest/largest selectivity in one trace (e.g., -O3), we further study the selectivity rankings of their correspondences in the other trace (e.g., -O0). The results are visualized in Fig.~\ref{fig:branch-dist}. Observe that in more than 98\% cases, they have the top-3 smallest or largest selectivity in the other trace.} %

Furthermore, we select 80 most challenging functions in Coreutils to further study the effectiveness of our path sampling strategy. 
These functions have more than 150 basic blocks and the average connectivity is larger than 3, namely, a block is connected to more than 3 blocks on average.
We compare the performance of 3 path sampling strategies.
The results are shown in Table~\ref{tab:path-sampling}. The three rows show the PR@1, the code coverage for -O0 and -O3 functions, respectively. The second column presents a strategy in which \toolname flips the last predicate encountered in the previous round with an uncovered branch. 
The third column denotes a strategy in which \toolname deterministically flips the predicates with the largest and the smallest selectivity
at each round. 
The last column presents our probabilistic path sampling strategy.
Observe that the probabilistic strategy substantially outperforms the other two and both the deterministic and probabilistic strategies can achieve good coverage. 

\input{tab_tex/tab_eva_path_sampling.tex}
\input{tab_tex/tab_eva_budget.tex}
\input{tab_tex/tab_eva_memory_models.tex}

\noindent
{\bf Code Coverage versus Precision.}  %
We run \toolname with different round budgets on  Coreutils and observe coverage and precision changes. The results are shown in Table~\ref{tab:code-coverage}.
Observe that if we only interpret each function once without any flipping, the precision is as low as 70 and the coverage is low too. 
With more budgets, namely, flipping more predicates, both the precision and the coverage improve, indicating \toolname can expose equivalent semantics. But the improvement becomes marginal after 200.  
\input{fig_tex/fig_eva_branch_relation.tex}
\input{fig_tex/fig_eva_branch_dist.tex}

\noindent
{\bf Probabilistic Memory Model~(PMM).} 
\textls[-6]{We run \toolname with different memory model setups on Coreutils to illustrate the benefit of modeling invalid memory accesses. 
The results are in Table~\ref{tab:pmm}. 
Specifically, {\it No-Mem} means we do not model invalid memory accesses. We return random values for invalid reads and simply discard invalid writes. 
The precision of No-Mem is nearly 10\% lower than PMM, while their coverage is similar.
That is because some dependencies between memory accesses are missing without handling invalid writes. On the other hand, if we allow writes to invalid memory regions but always return a constant value for all invalid reads, as shown in the column of {\it Const}, the precision is better than No-Mem. However, it is still inferior to PMM. This is due to returning the constant value making reads from different invalid addresses indistinguishable.}

\noindent
{\bf Robustness.} We alter system configurations of \toolname and run random sampling for each probabilistic component in \toolname. The experimental results show that \toolname is robust with regard to different configurations and variances in samplings. Details can be found in Section~\ref{supp:robu} of the appendix.\looseness=-1
}

%% file: tab_tex/tab_eva_baseline.tex
\newcommand{\problem}[1]{{\bf \color{red}#1}}
\definecolor{Gray}{gray}{0.85}
\newcolumntype{g}{>{\columncolor{Gray}}c}

\begin{table}[t]
\footnotesize
\setlength{\tabcolsep}{3.5pt}
\caption{Comparison of \toolname, IMF, and BLEX. \textbf{\textit{C}} and \textbf{\textit{G}} denote Clang and GCC, respectively. Each precision is averaged over the 106 binaries in Coreutils.
}
\label{tab:baseline1}
\centering
\revise{
    \begin{tabular}{cgccgcgc}
    \toprule
    \multirow{2.5}{*}{Pair} & \multicolumn{3}{c}{Precision@1} & \multicolumn{2}{c}{Precision@3} & \multicolumn{2}{c}{Precision@5}\\
         \cmidrule(lr){2-4} \cmidrule(lr){5-6} \cmidrule(lr){7-8}
         & \cellcolor{white}\toolname & IMF & BLEX & \cellcolor{white}\toolname & IMF & \cellcolor{white}\toolname & IMF\\
         \midrule
     \textbf{\textit{C}}-O0 \textbf{\textit{C}}-O3 
     & 94.5 & 77.5 & X & 98.2 & 84.2 & 98.7 & 86.4\\
     \textbf{\textit{C}}-O2 \textbf{\textit{C}}-O3 
     & 99.8 & 97.3 & X & 100.0 & 99.3 & 100.0 & 99.4\\
     \textbf{\textit{C}}-O0 \textbf{\textit{G}}-O3 
     & \textbf{94.5} & 60.1 & X & 97.3 & 70.6 & 98.6 & 73.4\\
     \textbf{\textit{G}}-O0 \textbf{\textit{G}}-O3 
     & 96.3 & 70.4 & 61.1 & 98.0 & 81.3 & 98.7 & 84.6\\
     \textbf{\textit{G}}-O2 \textbf{\textit{G}}-O3 
     & 98.6 & 89.5 & 77.1 & 99.4 & 95.5 & 99.8 & 96.2\\
     \textbf{\textit{G}}-O0 \textbf{\textit{C}}-O3 
     & \textbf{92.2} & 66.0 & X   & 94.1 & 76.1 & 96.2 & 80.0\\
     {\bf Average} & {\bf 96.0} & \textbf{76.8} & \textbf{69.1} & \textbf{97.8} & \textbf{84.5} & \textbf{98.7} & \textbf{86.7} \\ 
     \bottomrule
    \end{tabular}
    }
    \label{tab:my_label}
\end{table}

%% file: fig_tex/fig_eva_how_solve.tex
\begin{figure}[t]
    \centering
    \includegraphics[width=.98\linewidth]{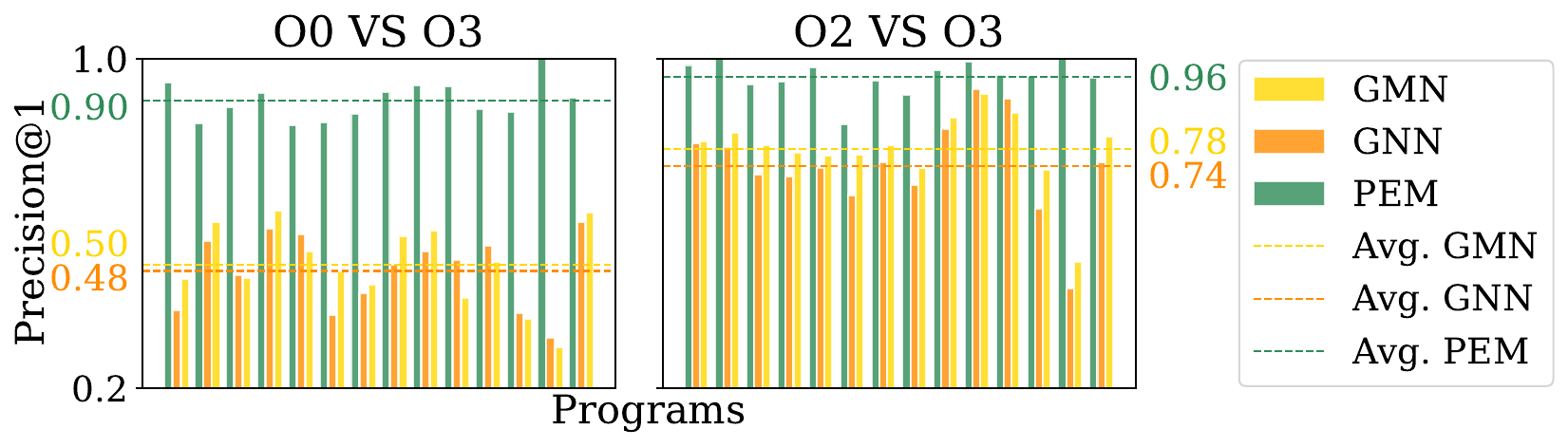}
    \vspace{-3pt}
    \caption{\textls[-20]{Comparison with How-Solve. We leverage the best two models (i.e., GNN and GMN) in How-Solve. Each bar denotes a program, whose name is elided. 
    A bar with 1.0 PR@1 means that \toolname finds the correct matches for all functions in the program. 
    Dashed lines denote the average PR@1 of each tool.}
    } 
    \label{fig:how-help}
    \vspace{-1pt}
\end{figure}

%% file: fig_tex/fig_eva_coverage.tex
\begin{figure}[t]
\vspace{-5pt}
    \centering
    \includegraphics[width=.43\textwidth]{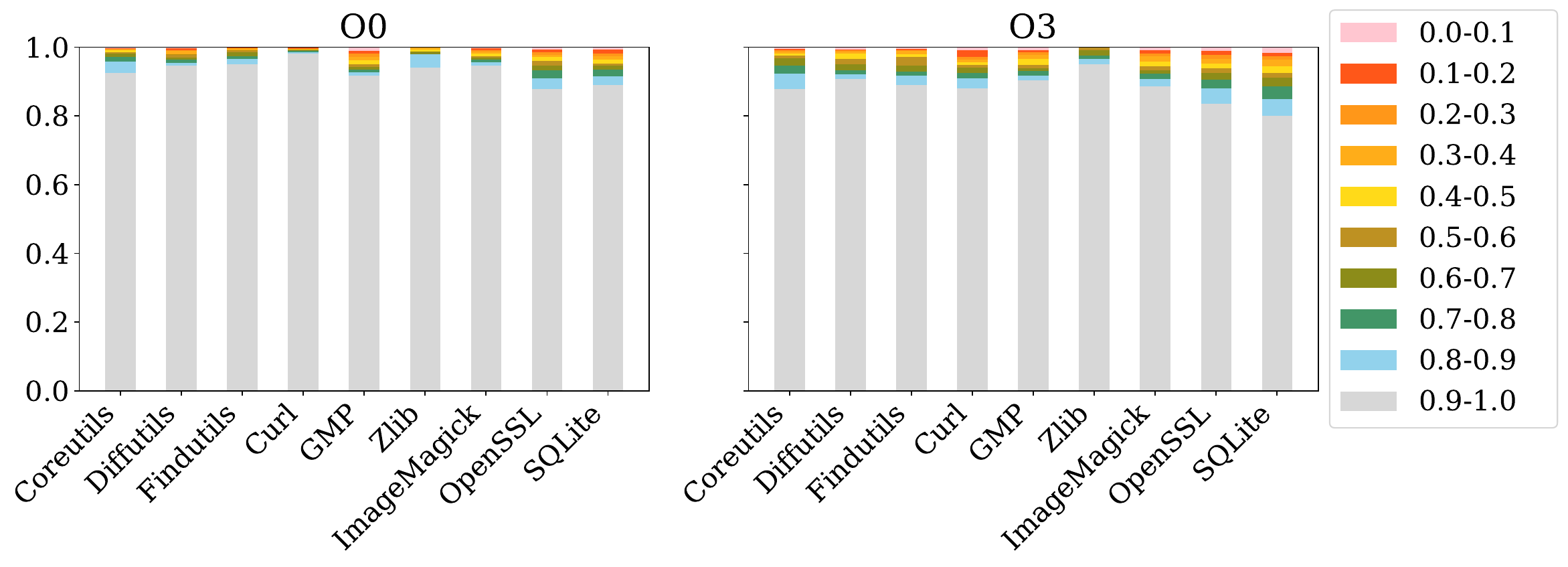}
    \vspace{-4pt}
    \caption{Coverage of \toolname 
    }
    \label{fig:Coverage}
    \vspace{-3pt}
\end{figure}

%% file: tab_tex/tab_eva_path_sampling.tex
\begin{table}[t]
\footnotesize
    \caption{\revise{Perf. w.r.t. Different Path Sampling Strategies}}
    \centering
    \revise{
    \begin{tabular}{rccc}
    \toprule
     & LastPred & Det. & \toolname \\ 
    \midrule
    PR@1 & 40.24  & 79.27 & {\bf 91.46}  \\
    Cover-O0 & 66.28   & 96.77 & 96.81  \\
    Cover-O3 & 53.14   & 92.95 & 92.97  \\
    \bottomrule
    \end{tabular}
    }
    \vspace{-14pt}
    \label{tab:path-sampling}
\end{table}

%% file: tab_tex/tab_eva_budget.tex
\begin{table}[t]
\footnotesize
    \caption{\revise{Perf. w.r.t. Different Budgets}
    }
    \centering
    \revise{
    \begin{tabular}{rccccccc}
    \toprule
     & 1 & 20 & 50 & 100 & 200 & 400 & 600\\ 
    \midrule
    PR@1        & 70    & 74    & 79    & 81    & 85    & 86    & 86 \\
    Cover-O0    & 63    & 87    & 91    & 94    & 96    & 98    & 98 \\
    Cover-O3    & 51    & 79    & 84    & 88    & 93    & 95    & 96 \\
    \bottomrule
    \end{tabular}
    }
    \vspace{-14pt}
    \label{tab:code-coverage}
\end{table}

%% file: tab_tex/tab_eva_memory_models.tex
\begin{table}[t!]
\footnotesize
    \caption{Perf. w.r.t. Different Memory Models}
    \centering
    \revise{
    \begin{tabular}{rccc}
    \toprule
     & No-Mem & Const & PMM \\ 
    \midrule
    PR@1        & 76.35    & 83.48    & {\bf 85.75}    \\
    Cover-O0    & 97.59    & 97.70    & {\bf 98.03}    \\
    Cover-O3    & 94.39   & 95.11    & {\bf 95.52}    \\
    \bottomrule
    \end{tabular}
    }
    \label{tab:pmm}
\end{table}

%% file: fig_tex/fig_eva_branch_relation.tex
\begin{figure}[t]
\vspace{-5pt}
    \centering
    \includegraphics[width=.31\textwidth]{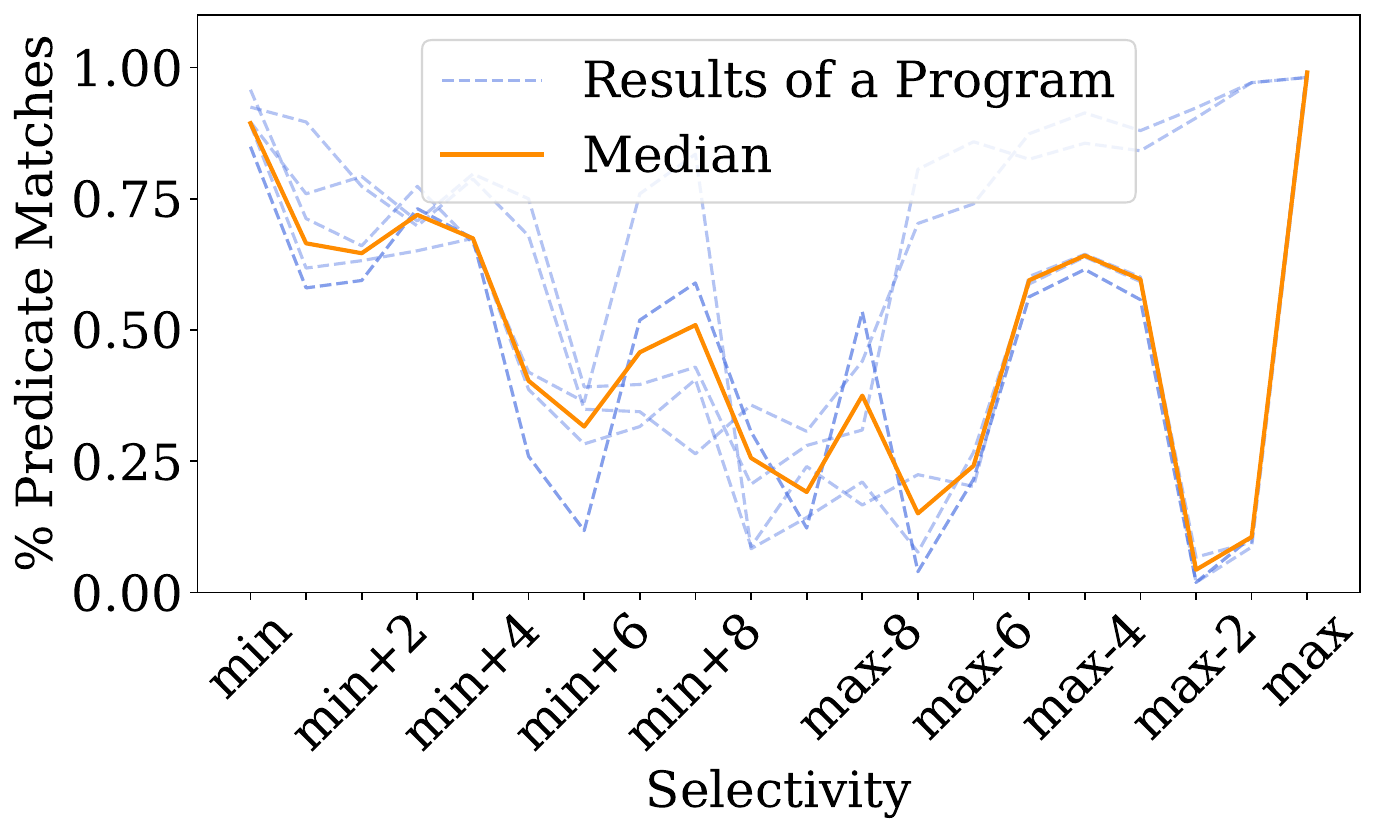}
    \vspace{-3pt}
    \caption{\revise{Predicate Correspondence versus Dynamic Selectivity.
    Each blue dashed line represents the analysis results of path pairs from two respective binaries compiled differently from a program.
The $x$ axis represents selectivity (with {\it min} the minimal and {\it max} the maximum) and $y$ denotes the percentage of predicate matches. 
We also compute the median for each selectivity, 
 resulting in the orange line.} 
}
    \label{fig:branch-relation}
    \vspace{-3pt}
\end{figure}

%% file: fig_tex/fig_eva_branch_dist.tex
\begin{figure}[t]
    \centering
    \includegraphics[width=.4\textwidth]{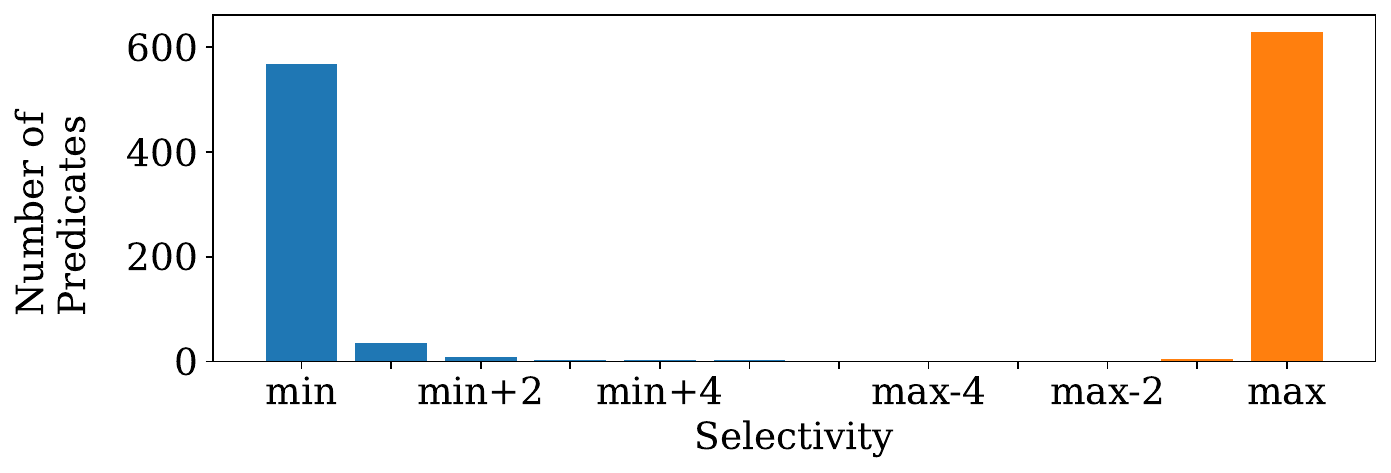}
    \vspace{-1pt}
    \caption{ \revise{\textls[-5]{Correspondence of Predicates with Min and Max Selectivity.
    Blue is for min and orange for max. For example, the bar at min+1 means that about 20 predicates with min selectivity in one trace have min+1 selectivity in the other~trace.} }
    }
    \label{fig:branch-dist}
    \vspace{-10pt}
\end{figure}

%% file: related.tex
\vspace{-5pt}
\section{Related Work}
\vspace{-3pt}
\noindent \textbf{Binary Similarity.} Many existing techniques aim to detect semantically similar functions, driven by static~\cite{graph,deepbindiff} and dynamic~\cite{bin-match,blex,IMF,vulseeker} analysis. A number of representative methods have been discussed in Section~\ref{sec:limitation}.
Other techniques compare code similarity at different granularity, e.g.,  whole binary~\cite{adiff,xu2021interpretation},  assembly~\cite{asm2vec,gmn,feng2016scalable}, and basic block~\cite{multimh}. 
While our method represents semantics at the function level, the resulting value sets of our system can be used as function semantic signatures and facilitate comparisons working at other granularity.

\noindent \textbf{Forced Execution.}
\textls[-3]{
Forced execution~\cite{blex,pmp,xforce,bda} concretely executes a binary along different paths by flipping branch outcomes. They typically aim to cover more code in a program and thus use coverage as the guidance. They can hardly select similar sets of paths for the same program compiled with different optimizations. 
Their focus is on recovering from invalid memory accesses. In contrast, the probabilistic memory model of \toolname reveals the different semantics introduced by different invalid accesses with high probability. }

%% file: suppl.tex
\setcounter{section}{0}

\section{$K$-Edge-Off Behavior Change across Optimization}
\label{supp:moti-example}

Take {\tt main\_cat} in Fig.~\ref{fig:moti_ex}  as an example. The condition check at line~\ref{line:cat:condi} is compiled to the structure shown in the blue circle of Fig.~\ref{fig:cfg:cat0}. The program first checks the value of {\tt flag0}. If {\tt flag0} is {\tt false}, it further checks the value of {\tt flag1}. If either evaluates to {\it true}, the program checks the value of {\tt format}. Note that {\tt format} always has the same value as {\tt flag0}, as shown at line~\ref{line:cat:f0}. Thus the check for {\tt format} is redundant when {\tt flag0} is {\it true}. In Fig.~\ref{fig:cfg:cat3}, the compiler removes this check from the {\it true}-path of {\tt flag0}. Assume that for some input the values of {\tt flag0}, {\tt flag1}, and {\tt format} are all {\it false}. Any path reaching {\tt simple\_cat} in Fig.~\ref{fig:cfg:cat0} has to be at least 2-edge-off since one has to flip either {\tt flag0} or {\tt flag1} and flip {\tt format}. In contrast, in Fig.~\ref{fig:cfg:cat3}, paths going to {\tt simple\_cat} are 1-edge-off by only flipping the branch at {\tt flag0} (due to optimization).
This suggests the 1-edge-off behavior of the two programs are quite different.

\input{suppl-method}

\input{suppl-theoretical}

\input{suppl-pmm.tex}

\input{suppl-impl}

\input{suppl-eval}

\clearpage
\input{tab_tex/tab_statistics}
\input{fig_tex/fig_time_eff.tex}

%% file: suppl-method.tex
\section{Logging Rules}
\label{supp:log-rule}

The logging rules are shown in Fig.~\ref{fig:supp:logging}. They only update the observable value statistics $OV$, without changing any other states. They have a higher priority than the interpretation rules, namely, they fire before the interpretation rules if both have the preconditions satisfied.
They only fire once for a unique combination of preconditions and states. 
This ensures values are properly logged (only once) before they are updated. The first three rules dictate that we record both address and value in memory accesses including those to the probabilistic memory.
\toolname also records predicates (Rule {\it LogCC}) and jump targets (Rules {\it LogJN} and {\it LogJR}).
Note that although different binaries from the same source may have different numeric values for comparisons and jump targets, \toolname normalizes these values to reveal similarity between equivalent programs. Details are discussed in Section~\ref{supp:impl} of this material.

%% file: suppl-theoretical.tex
\section{Advantages of Probabilistic Path Sampling}
\label{supp:theoretical}

\begin{figure}
    \centering
        \includegraphics[width=.6\linewidth]{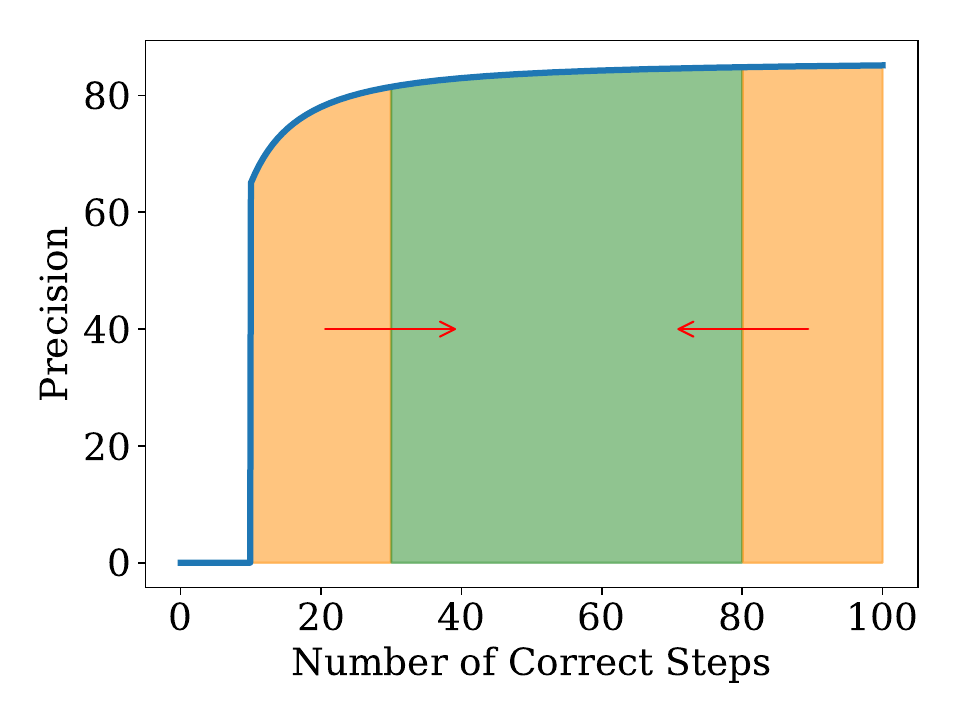}    
    
    \caption{Effects of the Probabilistic Path Sampling Algorithm. The $x$ axis denotes the number of correct steps. The $y$ axis denotes the PR@1 \toolname achieves. The dark blue line depicts the PR@1 of \toolname w.r.t. the number of correct steps. We show the effects of the probabilistic path sampling algorithms via red arrows. The algorithm moves cases from yellow areas to the green area.}
    \label{fig:step-precision}
\end{figure}

In this section, we prove that the probabilistic path sampling strategy outperforms the deterministic one.
We formally model the path selection workflow of \toolname as follows. Given a query program, \toolname first selects a set of paths to interpret and produces a set of observable value traces. Then, it applies the same path selection strategy to the pool of candidate programs which may include multiple versions of the query program. These versions are similar to the query program and the others are dissimilar. 
During the path selection for the programs in the pool, we say a predicate selection is correct {\em if the chosen predicate's correspondence in other versions of the program also has the smallest selectivity}\footnote{We only focus on proving the case of smallest selectivity as the other end is symmetric.}.
Intuitively, 
if \toolname makes more correct steps, it has better precision in the downstream similarity analysis. On the other hand, the return of growing correct steps becomes marginal when the number of correct steps becomes large, as the analysis already has sufficient information to expose similarity.
We hence hypothesize the relation between the number of correct steps and the similarity analysis precision follows a distribution like in Fig.~\ref{fig:step-precision} (and we will empirically demonstrate it later).

Our proof is hence to show that the probabilistic selection strategy can increase the density of the green area in Fig.~\ref{fig:step-precision} (denoting a middle range of correct steps),  with the cost of reduced density in the two orange areas on the two sides, when compared to the deterministic selection. 
Intuitively, having a higher density in the green area than the orange area to its left means that the probabilistic selection strategy can bias towards having a higher precision (by having a larger number of correct steps); having a higher density in the green area than the orange area to its right means that we sacrifice the chance of having a large number of correct steps, which is affordable because its gains on precision is marginal.

\input{fig_tex/fig_logging_rules.tex}

To formally prove it, we have the following steps. (1) We first empirically show that the relation between precision and number of correct steps is indeed like Fig.~\ref{fig:step-precision}, which can be approximated %
by a Pareto Distribution~\cite{arnold1983pareto}. (2) We further prove that the probabilistic selection strategy can improve the density for the mid-range correct steps. (3) At the end, we show that with the new density function and the Pareto Distribution, the expected precision is improved. 

\noindent
{\bf Step (1): Modeling Relation between Precision and Number of Correct Steps.}
It is difficult to compute the number of correct steps during the execution of \toolname. We hence approximate it using a simplified probabilistic model and derive the relation between precision and the {\em expected} number correct steps.

We first consider 
a {\it deterministic} path sampling strategy that exclusively flips the predicate with the smallest selectivity {\em in a single path}.
We use a pair of number $(c,e)$ to denote the state of \toolname at each step with $c$~(correct) denoting the number of previous steps that \toolname correctly selects a predicate,
and $e$~(error) denotes the number of previous steps that \toolname misses true correspondence.
At a given system state $(c,e)$, we define the probability that \toolname selects the next predicate correctly (from the aggregated pool of paths and predicates from all the previous steps) %
as follows.
\begin{equation}
 \label{eqt:pc}
  \mathcal{PC}(c, e) = \frac{c}{c+e}\times p_0
\end{equation}
Intuitively, it requires that \toolname first picks a path from a correct previous step (with  the probability of $\frac{c}{c+e}$) for further flipping and then selects the correct predicate to flip in this path (with a probability $p_0$).

The probability that a certain system state $(c,e)$ appears, denoted as $\mathcal{P}(c,e)$, can be calculated by the following:
\begin{equation}
\label{eqt:pce}
    \mathcal{P}(c,e) = \begin{cases}
    0 & c<0 \lor e<0 \\
    p_0 & c=1 \land e=0 \\
    1-p_0 & c=0 \land e \geq 1 \\
    \begin{minipage}{.2\textwidth}
    \footnotesize
    $\mathcal{PC}(c-1, e)\times \mathcal{P}(c-1,e) \\+ \Big(1-\mathcal{PC}(c, e-1)\Big ) \times \mathcal{P}(c, e-1)$
    \end{minipage}
     & Otherwise\\
    \end{cases}
\end{equation}

There are three initial cases in the formula.  The first case means $c$ and $e$ can never be less than $0$. The following two cases are related to the two outcomes of the first round of interpretation. 
For other steps, state $(c,e)$ is reached by 
either selecting a correct predicate at state $(c-1,e)$ or selecting a wrong predicate at state $(c,e-1)$. 
For a given budget $B$, $\mathcal{P}(i, B-i)$ denotes the probability that \toolname makes correct selections in $i$ steps. We treat $i$ as a random variable and study how it impacts the final precision (of function similarity analysis).

\begin{figure}[t]
    \centering
    \includegraphics[width=.6\linewidth]{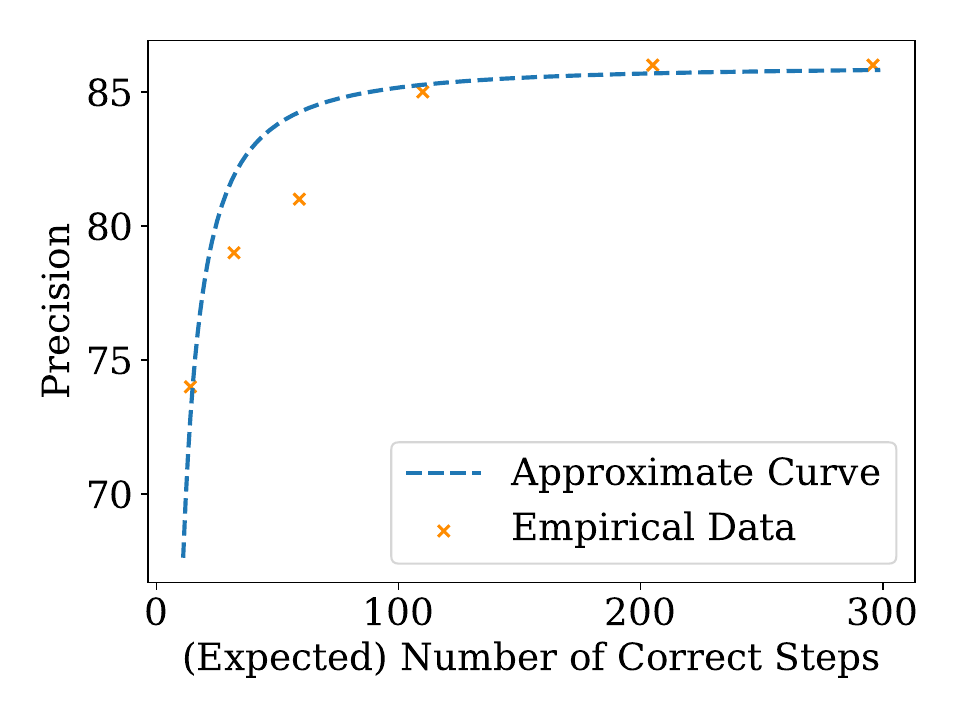}
    \caption{Number of Steps with Correct Predicate Selection w.r.t. Final Function Similarity Analysis Precision. The $x$ axis denotes the expected number of correct steps. The $y$ axis denotes the PR@1 \toolname achieves with the related number of correct steps.
    }
    \label{fig:emp-dist}
\end{figure}

We use 970 Coreutils functions and run 6 experiments with different budgets. While it is hard to quantify the number of correct steps in the real-world system, we use the probability mass function Equation~(\ref{eqt:pce}) to compute the expected number of correct steps for each budget. For example, if the budget is 400, we compute the expected number of correct steps via $\Sigma^{400}_{i=0} \mathcal{P}(i, 400-i)\times i$.
The (empirical) precision and the corresponding (expected) correct steps are shown in Fig.~\ref{fig:emp-dist}. 
The blue dashed curve tries to fit the data points. 
Observe that the precision increases sharply at the beginning with the increase of correct steps and
then the gains become marginal, which aligns well with Fig.~\ref{fig:step-precision}. 
Inspired by the Pareto Distribution~\cite{arnold1983pareto}, whose probabilistic density function has the aforementioned property, we use the following formula to model the relation between PR@1 and
the number of correct steps $i$  as follows.

\newcommand{\prmath}{\ensuremath{\mathcal{PR}_1}\xspace}

\begin{equation}
    \prmath(i)= \begin{cases}
    r_0 +\delta \times (1-(\frac{i_{min}}{i})^\alpha) & i > i_{min} \\
    0 & Otherwise\\
    \end{cases}
\end{equation}
In the above equation, $i_{min}$ is the minimal number of correct steps required to achieve a reasonable precision. The precision achieved with $i_{min}$ number of correct steps is denoted by $r_0$; $\delta$ and $\alpha$ are two parameters defining the max value of the precision 
and how fast the amount of increment decays as the number of correct steps grows. The blue curve in Fig.~\ref{fig:emp-dist} has the parameters $r_0=65, i_{min}=10, \delta=21, \alpha=1.4$.

\smallskip
\noindent
{\bf Step (2): Probabilistic Path Selection Improves Mid-range Correct Steps Density.}
The essence of probabilistic path selection is to have additional sampling at each step, flipping predicates with selectivity close to the smallest/largest.
Suppose that we spend $K$ steps from the budget to make additional sampling. The probability that \toolname selects a correct predicate at each step is updated as follows.
\begin{equation}
    \widetilde{\mathcal{PC}}(c, e)= \frac{c}{c+e}\times(p_0 + \frac{K}{B-K} \times \phi \times (1-p_0))
\end{equation}

$B$ is the budget, and $\phi$ is a factor representing the probability that \toolname indeed selects a correct predicate not having the smallest selectivity. Intuitively, when the correct predicate does not have the smallest selectivity (with the probability of $1-p_0$), \toolname has $\phi$ likelihood selecting the predicate when given an additional round. For example, the predicate likely has the next-to-smallest selectivity.
Note that we assume $K$ is relatively small to $B$. Thus we can ignore the noise introduced by additional samplings.

\begin{figure}[t]
    \centering
    \includegraphics[width=.7\linewidth]{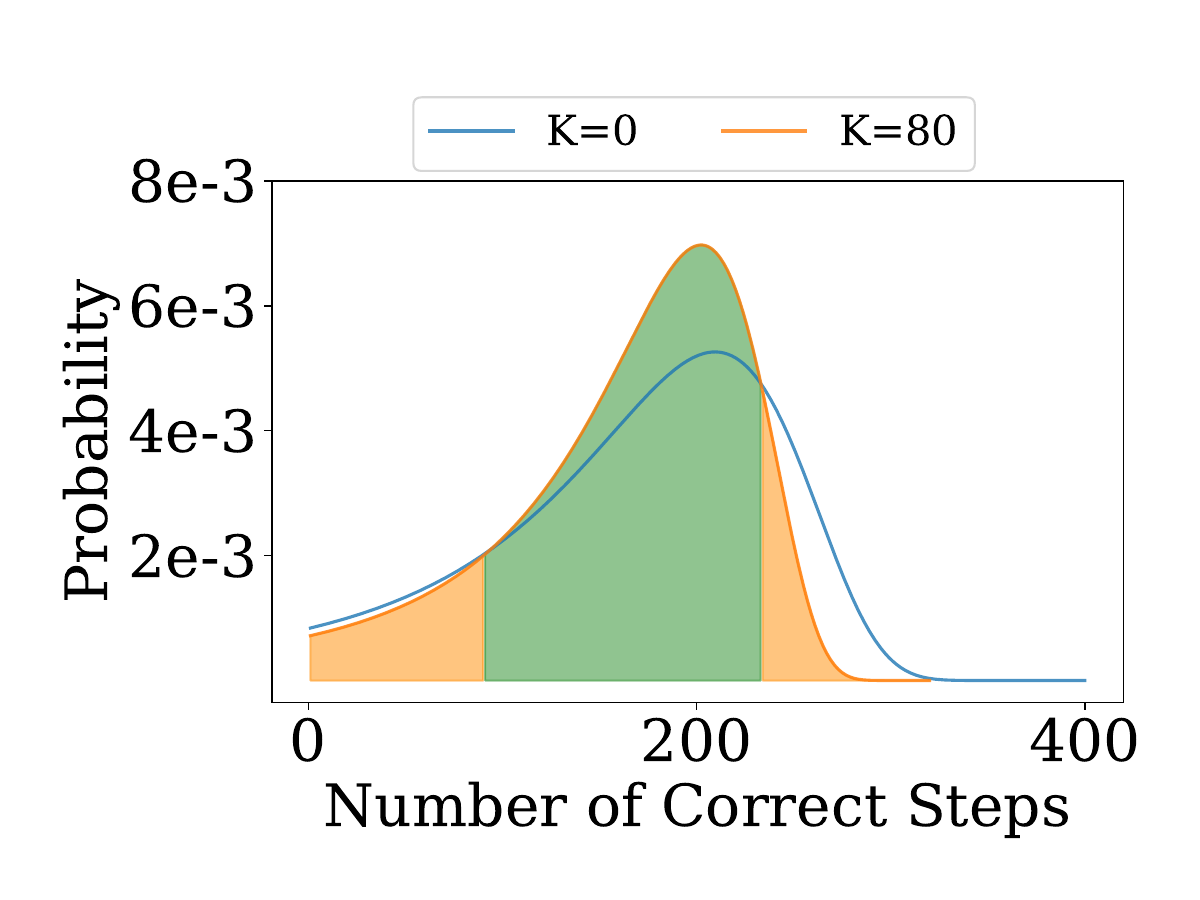}    
    \vspace{-2pt}
    \caption{Distributions of the Number of Correct Step. The $x$ axis denotes the expected number of correct steps. The $y$ axis denotes the probability. Blue and orange lines are for $K=0$ and $K=80$, respectively.
    Green and orange areas denote where the probabilistic sampling algorithm has a larger and smaller probability than the deterministic one, respectively.
    }
    \label{fig:effect-k-dist}
\end{figure}

We visualize the distribution difference introduced by the probabilistic sampling algorithm in Fig.~\ref{fig:effect-k-dist}. 
We can see that when $K=80$, \toolname has a larger probability to make 70--230 correct steps (i.e., the denser green area).

\smallskip
\noindent
{\bf Step (3): Probabilistic Path Selection Yields Better Expected Precision.}
Then we show that the overall expected PR@1, noted as $\mathbf{E}(\prmath(i))$, is increased. It is computed as follows.
\begin{equation}
    \mathbf{E}(\prmath(i)) = \sum^{B-K}_{i=0}\widetilde{\mathcal{PC}}(i, B-K-i)\times \prmath(i)
\end{equation}
The possible number of correct steps \toolname makes in total ranges from $0$ to $B-K$ (i.e., the total number of steps). For each possible number $i$, $\widetilde{\mathcal{PC}}(i, B-K-i)$ denotes the probability that \toolname makes $i$ correct steps in total, and $\prmath(i)$ denotes the expected PR@1 when \toolname makes $i$ correct steps.
We show in Fig.~\ref{fig:effect-k-exp} the distribution of $\mathbf{E}(\prmath(i))$ w.r.t. different values of $K$. We use $B=400, p_0=0.85$, and $\phi=0.75$. $K=0$ is equivalent to the deterministic selection. We can see that as $K$ grows, the expected PR@1 of \toolname improves. Thus the probabilistic sampling algorithm achieves a better performance than a deterministic one.

\section{Effect of Path Infeasibility}
In the following, we formally analyze two important properties of \toolname{} related to infeasible paths.

\subsection{Selecting Infeasible Paths Does Not Affect the Effectiveness of \toolname}
\label{appdx:proof-path-1}
Assume two equivalent programs $p$ and $p'$. We want to prove that \toolname{} can faithfully disclose their equivalence even when it selects infeasible paths.
We first consider a simplified scenario in which each path in $p$, including infeasible path, has a corresponding path in $p'$, meaning that the two paths produce the same sequence of observable values. This precludes code removal type of optimizations. Starting from two seed paths, which are equivalent as they are derived from the same input (on the two equivalent programs), we can derive that the predicates having the smallest and largest selectivity correspond to each other along the two seed paths. 
As such, flipping them yields two equivalent new paths regardless of their feasibility.
This can be turned in a formal inductive proof. Details are elided.

Complexity arises when $p'$ is $p$ with certain (dead) code removed. In such cases, \toolname may select a path in $p$ that may include the removed code and does not have the correspondence in $p'$. We prove that by leveraging the observation that code removal is rare (due to the difficulty of proving path feasibility/infeasibility). As such, we assume any path in $p$ has a large probability $t$ to have an equivalent path in $p'$. We can hence derive a probabilistic proof similar to the above. Details are elided.

\subsection{Selecting Infeasible Paths Does Not Conclude Equivalence for Dissimilar Programs}
\label{appdx:proof-path-2}

\input{fig_tex/fig_dist_expectation.tex}
\input{fig_tex/fig_dist_path_sim.tex}

Assume $b_1$ and $b_2$ are an arbitrary pair of predicates from two different binary programs $p_1$ and $p_2$. Since $p_1$ and $p_2$ are different, we suppose that no more than half of the successors of $b_1$ are the same with $b_2$.
We prove that most infeasible paths that \toolname samples in $p_1$ will not coincide with paths in $p_2$.
The proof consists of three key steps.
First, we show that for any path sampled from $p_1$, the probability that it coincides with a path in $p_2$ decreases exponentially when it contains growing flipped predicates. Then we demonstrate that most paths sampled by \toolname contain at least 4 flipped predicates (given a sample budget of 400 paths). Finally, we show that the expected number of coincided paths is only 25, whose effects are limited out of the 400 sampled paths in total. 

Recall that we use the term $k$-edge-off path to denote a path where $k$ predicates on it are flipped.
We denote as $\mathcal{P}_{same}(k)$ the probability that a $k$-edge-off path from $p_1$ coincides with a path in $p_2$. An upper bound of $\mathcal{P}_{same}(k)$ is computed as follows.
\begin{equation}
\label{eqt:psame}
    \mathcal{P}_{same}(k) \leq 
    \begin{cases}
        1 & k=0\\
        \mathcal{P}_{same}(k-1) * 0.5 + \overline{\mathcal{P}_{same}(k-1)}*0 & k > 0\\
    \end{cases}
\end{equation}

The first row of Equation~\ref{eqt:psame} shows the probability of coincidence for a seed path (i.e., the faithful path of executing $p_1$ on seed values without flipping any predicate). We give a pessimistic estimation by assuming the seed path in $p_1$ is equivalent to a path in $p_2$ with a probability as high as  $1$. 

The second row of Equation~\ref{eqt:psame} shows the probability that a $k$-edge-off path in $p_1$ coincides with a path in $p_2$ for any $k>0$.
Note that a $k$-edge-off path is derived from a $(k-1)$-edge-off path. The first term hence depicts the case that the $(k-1)$-edge-off path coincides with a path in $p_2$. In the worst case, the predicate \toolname chooses to flip in $p_1$ has an aligned predicate on the corresponding path in $p_2$. 
Flipping the predicate has at most $0.5$ probability to lead the execution of a basic block that is in $p_2$, resulting in an equivalent $k$-edge-off path.
On the other hand, if the $(k-1)$-edge-off path already contains basic blocks that are not in $p_2$, the execution context will not be in alignment with any execution context in $p_2$. Thus the derived $k$-edge-off path will not coincide with any path in $p_2$.

We have shown that the probability of coincidence for a $k$-edge-off path exponentially decreases as $k$ grows. Next we reason about the distribution of $k$ by modeling the sampling behavior of \toolname. At each step, \toolname flips the predicate with the largest/smallest dynamic selectivity. 
We assume the probability is the same for each path that a selected predicate comes from the path.
The probability of the selected predicate being on a $k$-edge-off path is then $\frac{\#\ of\ \textit{k-edge-off\ paths}}{\#\ of\ all\ sampled\ paths}$. 

We use $\mathcal{N}(k, i)$ to denote the expected number of $k$-edge-off paths at the $i$-th step. It is calculated as follows.
\begin{equation}
\label{eqt:number-k-off}
\mathcal{N}(k, i)=
\begin{cases}
    1 & k=0 \land i\geq 1 \\
    0 & k>0 \land i=1 \\    
    \mathcal{N}(k, i-1) + \frac{\mathcal{N}(k-1, i-1)}{i-1} * 1 & k>0 \land i>1 
\end{cases}
\end{equation}
The first two lines of Equation~\ref{eqt:number-k-off} denotes that at the first step, \toolname faithfully samples a $0$-edge-off path without flipping any predicates.
The third line computes the number of $k$-edge-off paths as follows.
At step $i$, the expected number of $k$-edge-off paths is the sum between the number of $k$-edge-off paths at the previous step (i.e., $\mathcal{N}(k, i-1)$) and the probability that \toolname flips a predicate on a $(k-1)$-edge-off-path (i.e., $\frac{\mathcal{N}(k-1, i-1)}{i-1}$).
We visualize the distribution of $\mathcal{N}(k,400)$ (i.e., the number of $k$-edge-off paths after using up a sample budget of 400) in Fig.~\ref{fig:k-off-dist}. 

For each $k$, the expected number of coincided paths is calculated by 
$\mathcal{P}_{same}(k)\times${\it \#\ of k-edge-off paths}. Its distribution is depicted by the orange curve in Fig.~\ref{fig:k-off-dist}.
The expected number of coincided paths is 25 when \toolname samples 400 paths from $p_1$. They are a relatively small portion of all the sampled paths, and hence insufficient to induce misclassification in \toolname{}. $\Box$

\input{fig_tex/fig_dist_k.tex}

\subsection{Empirical Study}
\label{appdx:path-emp}
We further validate the above two proofs with an empirical study. We randomly sample 50 main functions from the Coreutils binary compiled with the {\tt -O0} option. Then we use their corresponding functions compiled with the {\tt -O3} option to construct 50 pairs of similar functions, and use another 50 randomly selected {\tt -O3} functions to construct 50 pairs of dissimilar functions. For each pair of functions, \toolname samples 400 paths from both functions, respectively. Then for each path from the {\tt -O0} function, we find the most similar path in the {\tt -O3} function and compute the similarity score. The results are visualized in Fig.~\ref{fig:path-sim}.

From Fig.~\ref{fig:path-sim-hist}, we can see that most paths from similar function pairs match with significantly higher similarity scores. It demonstrates that \toolname indeed collects similar observable values from equivalent paths regardless of feasibility. The score is not 100\% mainly because we do not model all floating point instructions and library calls.
On the other hand, from Fig.~\ref{fig:path-sim-scatter}, we can see that for any pair of dissimilar functions, the number of paths matched with high similarity scores (e.g., scores $>0.6$) is significantly smaller than the number of those from the similar functions. It validates that \toolname does not sample too many similar paths from dissimilar function pairs that leads to wrong conclusions of similarity.

%% file: fig_tex/fig_logging_rules.tex
\begin{figure}[t]
\begin{mdframed}
\begin{minipage}{\textwidth}
{\footnotesize
\begin{mathpar}
\inferrule{Regs[r_2]=a\\ \textbf{valid}(a) \\ v=M[a]\\ OV'=\update{OV}{a}{OV[a]+1}\\
OV''=\update{OV'}{v}{OV'[v]+1}\\
}
{\progstate{Regs}{r_1 = [r_2];I}{ic}{M}{OV}
\\ \\ \\ \\ \rightarrow
\progstate{Regs}{r_1 = [r_2];I}{ic}{M}{OV''}
}
{LogLd}\\
\inferrule{Regs[r_2]=a\\ \neg\textbf{valid}(a) \\  v=\textbf{invalidLd}(a)\\ \\ \\ \\ OV'=\update{OV}{a}{OV[a]+1}\\ \\ \\ \\ \\ \\ \\ \\
OV''=\update{OV'}{v}{OV'[v]+1}\\
}
{\progstate{Regs}{r_1 = [r_2];I}{ic}{M}{OV}
\\ \\ \\ \\ \rightarrow
\progstate{Regs}{r_1 = [r_2];I}{ic}{M}{OV''}
}
{LogIvLd}\\
\inferrule{Regs[r_1]=a\\ Regs[r_2]=v\\ \\ \\ \\ \\ 
OV'=\update{OV}{a}{OV[a]+1}\\ \\ \\ \\ \\ \\ \\ 
OV''=\update{OV'}{v}{OV'[v]+1}
}
{\progstate{Regs}{[r_1] = r_2;I}{ic}{M}{OV}
 \rightarrow\\
\progstate{Regs}{[r_1] = r_2;I}{ic}{M}{OV''}
}
{LogSt}\\
\inferrule{
OV'=\update{OV}{a}{OV[a]+1}\\
}
{\progstate{Regs}{\langkw{jmp}\ a;I}{ic}{M}{OV}
\rightarrow\\ \\
\progstate{Regs}{\langkw{jmp}\  a;I}{ic}{M}{OV'}
}
{LogJN}\\
\inferrule{Regs[r] = a\\
OV'=\update{OV}{a}{OV[a]+1}\\
}
{\progstate{Regs}{\langkw{jr}\ r;I}{ic}{M}{OV}
\rightarrow
\progstate{Regs}{\langkw{jr}\ r;I}{ic}{M}{OV'}
}
{LogJR}\\
\inferrule{Regs[r_2]=v_2\\ Regs[r_3]=v_3\\ \\ \\ 
v_1 = \textbf{condValue}(\otimes, v_2, v_3)\\
OV'=\update{OV}{v_1}{OV[v_1]+1}\\
}
{\progstate{Regs}{r_1 = r_2 \otimes r_3;I}{ic}{M}{OV} 
\\ \\ \\ \rightarrow
\progstate{Regs}{r_1 = r_2 \otimes r_3;I}{ic}{M}{OV'}
}
{LogCC}
\end{mathpar} 
} 
\end{minipage}
\label{fig:sem:rule_sem}
\end{mdframed}    
\caption{Logging Rules}
\label{fig:supp:logging}
\end{figure}

%% file: fig_tex/fig_dist_expectation.tex
\begin{figure}[t]
    \centering
    \includegraphics[width=.6\linewidth]{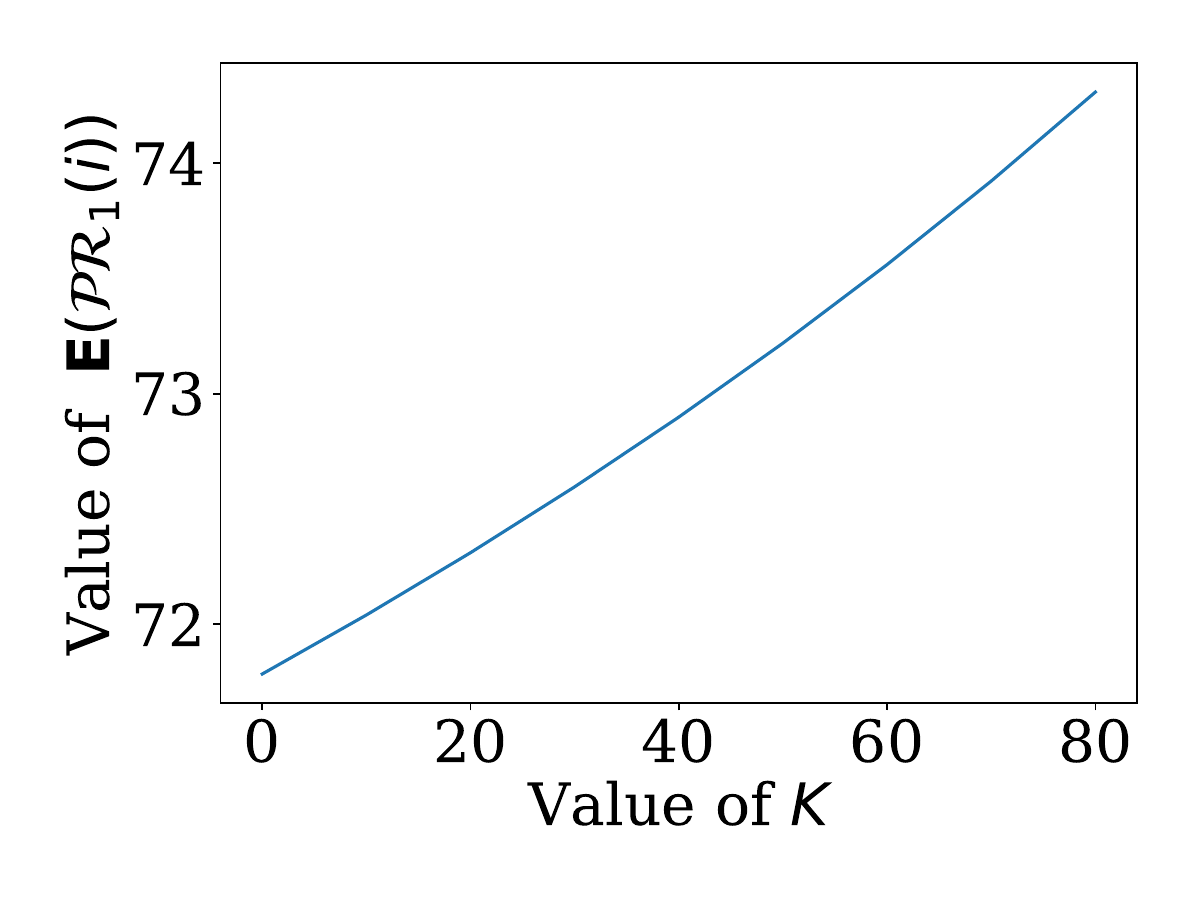}    
    \caption{$\mathbf{E}(\prmath(i))$ w.r.t. $K$. The $x$ axis denotes the value of $K$ (i.e., the total number of additional samplings). The $y$ axis denotes the value of $\mathbf{E}(\prmath(i))$ (i.e., the expectation of reward)}
    \label{fig:effect-k-exp}

    
\end{figure}

%% file: fig_tex/fig_dist_path_sim.tex
\begin{figure*}[t]
\begin{subfigure}[t]{.4\textwidth}
    \includegraphics[width=\textwidth]{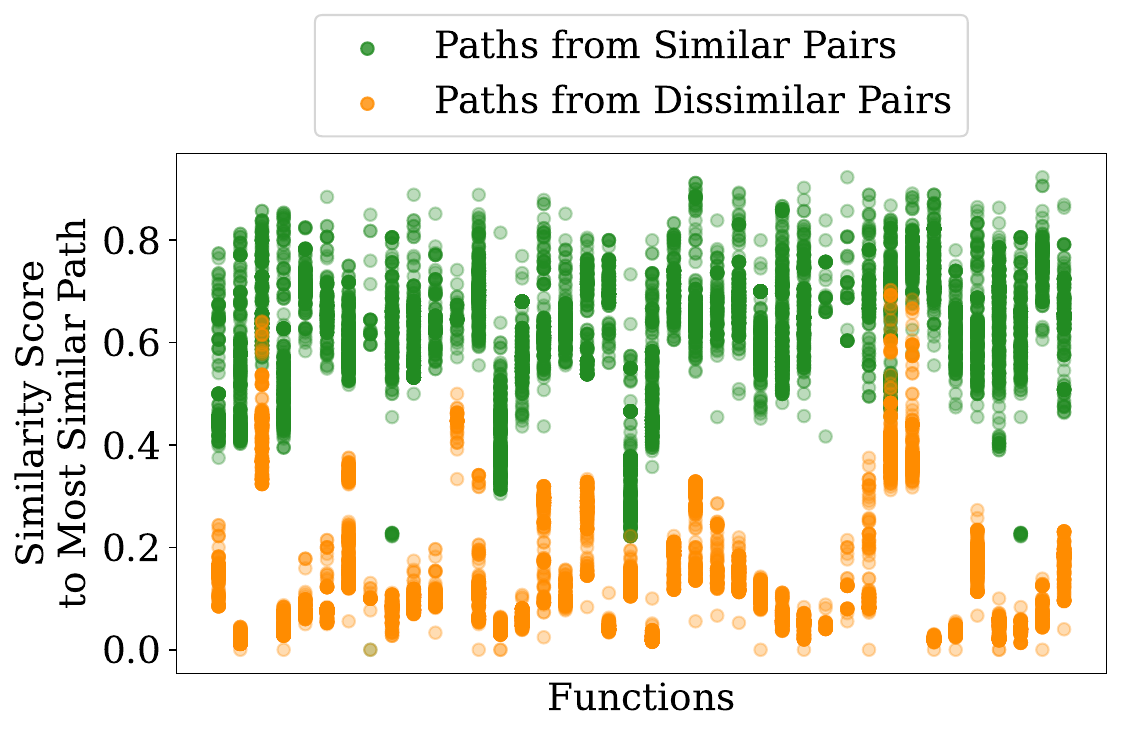}
    \caption{The $x$-axis denotes functions, whose names are elided. The $y$-axis denotes the similarity score of paths in the {\tt -O0} function to the most similar path in the {\tt -O3} function. Green and orange points denote paths from similar and dissimilar pairs.}
    \label{fig:path-sim-scatter}
\end{subfigure}
\hspace{10pt}
\begin{subfigure}[t]{.4\textwidth}
    \includegraphics[width=\textwidth]{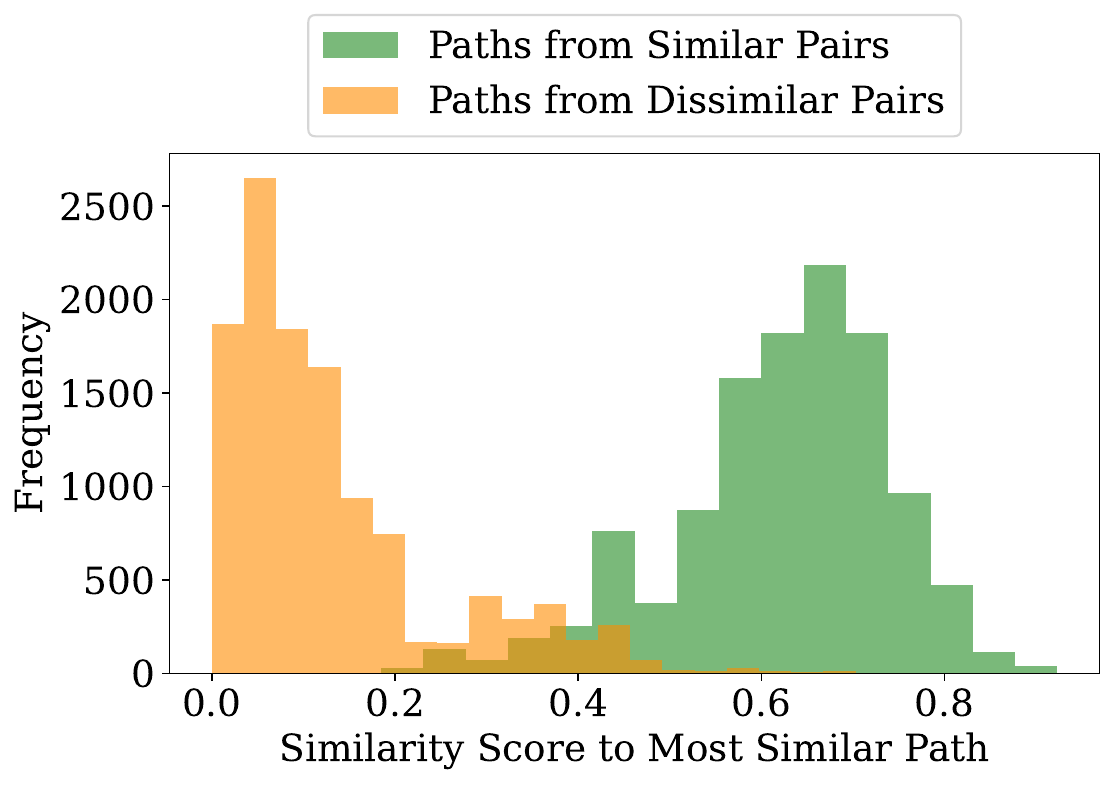}
    \caption{The $x$-axis denotes the similarity score of paths from the {\tt -O0} function to the most similar path from the {\tt -O3} function. The $y$-axis denotes the frequency of paths. Green and orange bars denote distributions of similar and dissimilar pairs.}
    \label{fig:path-sim-hist}
\end{subfigure}
    \caption{Distribution of Path Similarity. Fig.~\ref{fig:path-sim-scatter} shows
    paths from similar and dissimilar pairs w.r.t. functions; Fig.~\ref{fig:path-sim-hist} aggregates the distribution of  similarity scores.}
    \label{fig:path-sim}    
    \vspace{-2pt}
\end{figure*}

%% file: fig_tex/fig_dist_k.tex
\begin{figure}[t]
    \centering
    \includegraphics[width=.8\linewidth]{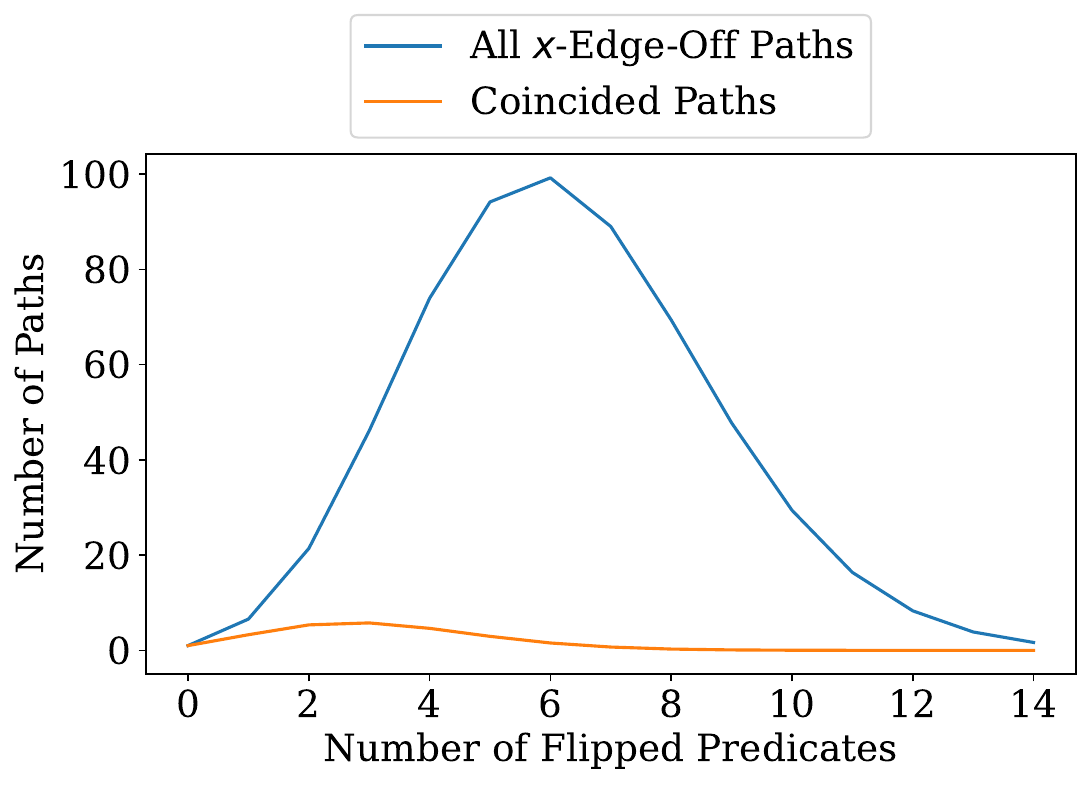}
    \caption{Distribution of $k$. The $x$-axis denotes the number of flipped predicates on a path. The $y$-axis denotes the number of paths. The blue curve depicts the number of all $x$-edge-off paths. The orange curve denotes the number of coincided paths. It is obtained by multiplying each point on the blue curve with $\mathcal{P}_{same}(x)$.
    }
    \label{fig:k-off-dist}
\end{figure}

%% file: suppl-pmm.tex
\section{Example for Probabilistic Memory Model}
\label{supp:example-pmm}

\smallskip
\noindent
{\bf Example.}
Consider the two functions {\it findDouble} and {\it findPoint} in Fig.~\ref{fig:mem-model}. The former traverses a linked list with a node size 0x10 and the latter traverses a linked list with a node size 0x18. 
Fig.~\ref{fig:PM-region} shows a PM by \toolname, which has a size of 128 and is filled with random values. Assume in the interpretations of both functions, the pointer to the root of the respective linked-list is null (e.g., the preceding linked-list allocation and initialization are bypassed due to predicate flipping). With our PMM, both functions access the invalid address $0x00$ and get the same value (because our PMM is equivalence preserving). However, as the interpretation progresses, these two functions access different sets of addresses. For {\it findDouble}, it treats the value at address $0x08$ as the pointer to the next node. When it tries to access the address $0x33...5418$ (the value stored at $0x08$), our PMM maps it to an address $0x18$ in PM via the operation {\tt mod 128}. Then it reads from $0x18$ and treats the value at $0x20$ as the next pointer, and so on. The cells accessed by {\it findDouble}, $0x00-0x18-0x50$, are chained by the blue arrows in the figure. In contrast, although {\it findPoint} also accesses $0x00$ at the beginning, the pointer field of the node it traverses is at $0x10$, resulting in a different access chain $0x00-0x30-0x60$, connected by the red arrows. As such, the observable values are different for the two functions, denoting their different semantics. 
In addition, one can easily tell that two versions of {\it findDouble}, optimized and unoptimized, would produce the same sequence of observable values. 
Note that if the two functions traverse a linked list of the same node size, their behaviors are indeed not separable. 
$\Box$
\input{fig_tex/fig_mem_model.tex}
\input{fig_tex/fig_pm_region.tex}

%% file: fig_tex/fig_mem_model.tex
\begin{figure}[t]
\begin{minipage}{\textwidth}
\begin{lstlisting}[deletekeywords={input}, style=customc, firstnumber=1, xleftmargin=0.5cm, basicstyle=\scriptsize,
    escapeinside={(|}{|)}]
struct DoubleList{
    /* offset 0x0 */ double val;
    /* offset 0x8 */ DoubleList* next;};
struct PointList{
    /* offset 0x0  */ int64 x,y;
    /* offset 0x10 */ PointList* next;};
DoubleList* findDouble(DoubleList* list, double val){ (|\label{line:ll:func}|)
    while(list){
        if(val == list->val) (|\label{line:ll:val}|)
            return list;
        list = list->next; (|\label{line:ll:next}|)
    }
    return list;
}
PointList* findPoint(PointList* list, int64 x, int64 y){ (|\label{line:ll:func}|)
    while(list){
        ...
        list = list->next; (|\label{line:ll:next}|)
    } ...
}
\end{lstlisting}
\end{minipage}
\vspace{-5pt}
\caption{Example Illustrating Probabilistic Memory Model}
\label{fig:mem-model}
\end{figure}

%% file: fig_tex/fig_pm_region.tex
\begin{figure}[t]
    \centering
    \includegraphics[width=.40\textwidth]{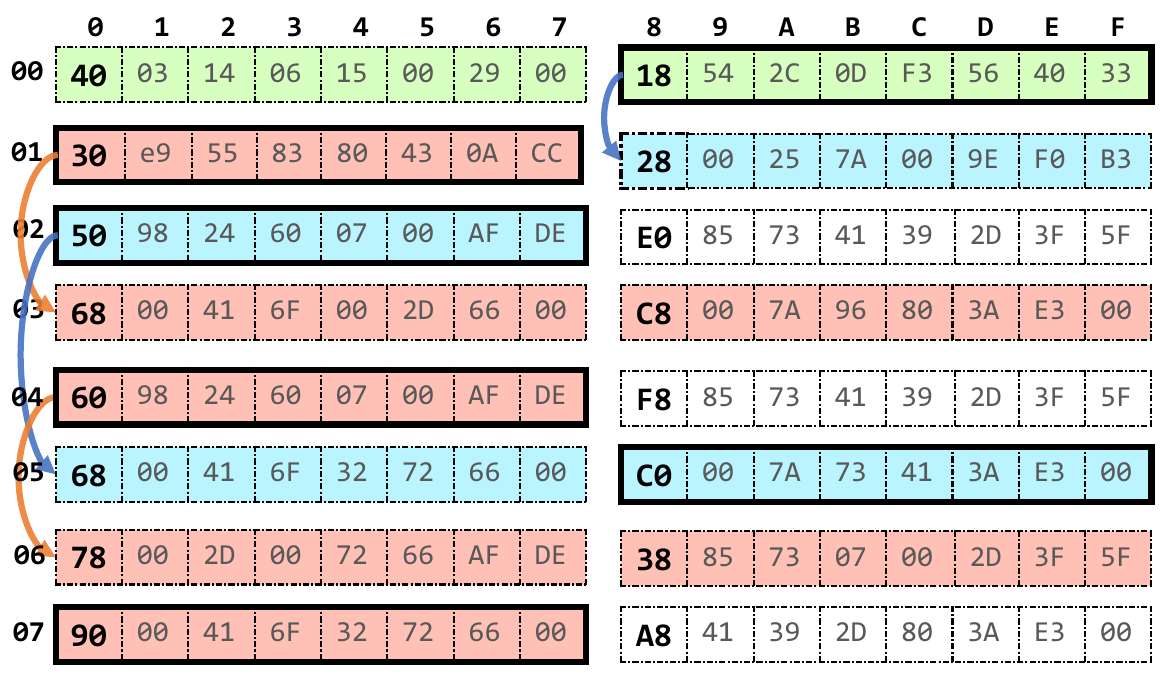}
    \caption{Example of Probabilistic Memory Region. The numbers on the top denote the lowest 4 bits of memory addresses, and the numbers on the left denote the highest 8 bits. Cells with bold texts are the least significant byte (LSB). 
    Memory cells in blue are exclusively accessed by {\tt findDouble} and those in red are exclusively accessed by {\tt findPoint}. Green cells are accessed by both functions and white cells are accessed by neither. Also, cells in bold boxes
    are interpreted as pointers, with arrows pointing to their targets.}
    \label{fig:PM-region}
\end{figure}

%% file: suppl-impl.tex
\revise{

\section{Implementation}
\label{supp:impl}

As shown in Fig.~\ref{fig:wf}, \toolname consists of two key components: 
a {\it probabilistic execution engine} (in the left grey box) that interprets binary programs and collect the observable values, and a {\it value analyzer} (in the right grey region) that aggregates sets of observable values collected from different paths and normalizes values according to their type.

\smallskip
\noindent
{\bf Probabilistic Execution Engine.} 
We use IDA~\cite{idapro} and Ghidra~\cite{ghidra} to obtain static information about binary programs, and build the probabilistic execution engine on top of a well-known emulator QEMU~\cite{qemu}. QEMU translates binary programs to an intermediate representation (IR) that is similar to the language defined in Fig.~\ref{fig:definition}. We then implement our semantic rules (Fig.~\ref{fig:sem_rule}) and logging rules (Fig.~\ref{fig:supp:logging}) on the QEMU-IR. To reduce invalid memory accesses and provide more realistic execution contexts, we also model the local-stack for each function and  the dynamically allocated heap memory.
Recall that for a binary program, \toolname interprets it for multiple rounds with a predicate flipped in each round (see Algorithm~\ref{alg:path-sampling}). The flipped predicates are selected following the probability density function~(PDF) of a Beta-distribution with the shape parameter $\alpha=\beta=0.03$, noted as $\mathcal{B}_{003}$.  $\mathcal{B}_{003}$ has a U-shape with large values near $0$ and $1$ and small values in the middle. To select the predicate to flip from a list of candidates, \toolname first samples a random number $i \sim \mathcal{B}_{003}$. Then we normalize the range of $i$ from $(0,1)$ to $[0,1]$ via the following formula:
\begin{equation*}
    i' = clip(clip(i-0.05, 0, 1)\times\frac{10}{9},0,1)
\end{equation*}
Then \toolname sorts the list of candidates by their dynamic selectivity and then selects the predicate as follow:
\begin{equation*}
    \texttt{candidates}[i' \times (len(\texttt{candidates})-1)]
\end{equation*}

\smallskip
\noindent
{\bf Value Analyzer.}
The value analyzer first aggregates the observable values $OV$ by adding up the number of observations in all paths for each value. Then it normalizes the values according to their types: for values that are {\it jump targets}, we leverage the dynamic linking information (these information are available even in stripped binary files) to look up external library functions that are potentially associated with the jump targets. If a jump target is not associated with external functions, we remove it from the observable value set. For values with the string type, we truncate the values at the string terminator (i.e., the value 0). For values collected in {\it predicate comparisons}, we compute their dynamic selectivity. All other values remain unchanged.
Finally, the value analyzer sorts values by their number of observations and only selects the top 50,000 most frequent values. For similarity comparison, we use the standard metric for set comparison, i.e., the Jaccard index~\cite{IMF}.

}

%% file: suppl-eval.tex
\section{Robustness of \toolname}
\label{supp:robu}

\smallskip
\noindent
{\bf Robustness w.r.t. System Configurations.} %
We alter the system configurations of \toolname and analyze how each component affects the performance. For each configuration, we run the experiment on Coreutils with 5 alternative values. The results are visualized in Fig.~\ref{fig:sys-abl}. We can see that the performance of \toolname does not change significantly in most cases, meaning that \toolname is robust when system configuration changes. In the setup that we only unroll loops for one time, the precision@1 and the coverage for both O0 and O3 binary drop a bit. That is because compilers usually conduct sophisticated optimizations to unroll loops, resulting in a large number of basic blocks in the loop body. Table~\ref{tab:config} lists the detailed results for all the configurations. Note that we also alter how \toolname initializes input values for functions, shown in the last 5 rows. {\it Same} means \toolname initializes all the input parameters with the same value; {\it Rand1-4} means \toolname initializes parameters at different positions with different random values and we conduct the experiments with 4 different random seeds. We can see that initializing parameters differently is slightly inferior than initializing parameters equivalently because the order of parameters may change among optimizations.  

\input{fig_tex/fig_sys_abl.tex}
\input{tab_tex/tab_config.tex}

\begin{figure}[t]
    \centering
    \includegraphics[width=.3\textwidth]{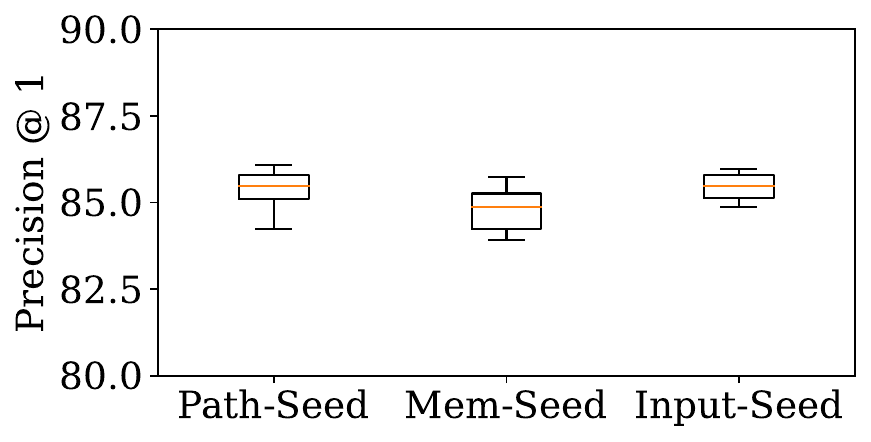}
    \caption{\revise{Perf. w.r.t. Different Random Seeds}
    }
    \label{fig:robu-prob}
\end{figure}

\smallskip
\noindent
{\bf Robustness w.r.t. Probabilistic Variance.} 
For each probabilistic component in \toolname, we run the related random sampling process with 10 different seed values. As shown in Fig.~\ref{fig:robu-prob}, \toolname is robust with regard to variances in samplings. 

\section{Performance Variation w.r.t. Negative Sample Ratios}
\label{supp:ratio-perf}
A dataset is composed of negative samples (dissimilar pairs) and positive samples (similar pairs). A class ratio N:1 means that in a query, apart from the ground truth target function, we also sample N other functions as the negative data. Existing machine-learning-based work samples at a relatively low ratio. For example, Trex samples at a ratio of 5:1 and achieves good precision. 
However, 5:1 is not realistic in real-world scenarios. A tool may need to compare a function with many other functions to find the real similar one. In our setup, there is only one true positive and all the other functions are true negatives. The ratio is usually very large. 
We argue a ratio around 1:1000 is more realistic.

We compute the PR@1 of \toolname, Trex and SAFE with different ratios including 1:1, 5:1, 10:1, 20:1, 50:1, 100:1, 500:1, and $\infty$:1. $\infty$ is the number of all functions in the project.
As Fig.~\ref{fig:baseline2-2} depicts,
while the PR@1 of all three tools decreases as the ratio increases, \toolname has more stable performance. For example, \toolname degrades 23\% (from 100\% to 77\%) on OpenSSL as the ratio increases from 1:1 to $\infty$:1, whereas the precision of the other two tools degrades to less than 20\%. This reveals that it is more challenging when the ratio becomes larger,
and \toolname is more resilient to class ratio changes and  likely to perform better in real-world scenarios.

\section{Searching for One-Day Vulnerabilities}
\label{supp:cve}

Detailed results are shown in Table~\ref{tab:cve}. \toolname lists the problematic function with the highest score in most cases. For the case that \toolname ranks it at 43, we manually check the function and find it delegates most logic to external library functions, thus \toolname does not find enough semantic information. The results can be improved by modeling the behavior of calling to external functions.

\input{fig_tex/fig_attention.tex}
\input{tab_tex/tab_cve}
\input{fig_tex/fig_baseline2-2}
\input{tab_tex/tab_multi_arch}

%% file: fig_tex/fig_sys_abl.tex
\begin{figure}[t]
    \centering    
    \begin{subfigure}[t]{.2\textwidth}
        \includegraphics[width=\textwidth]{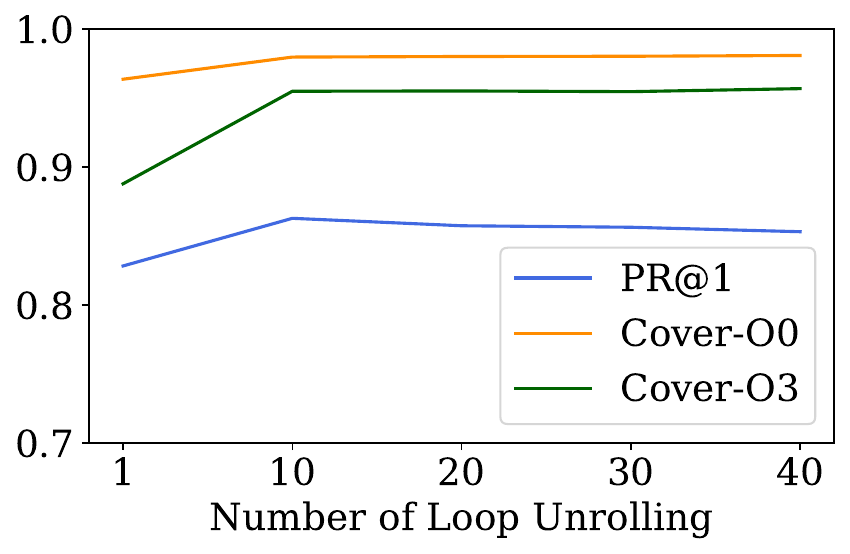}
    \end{subfigure}
    \begin{subfigure}[t]{.2\textwidth}
        \includegraphics[width=\textwidth]{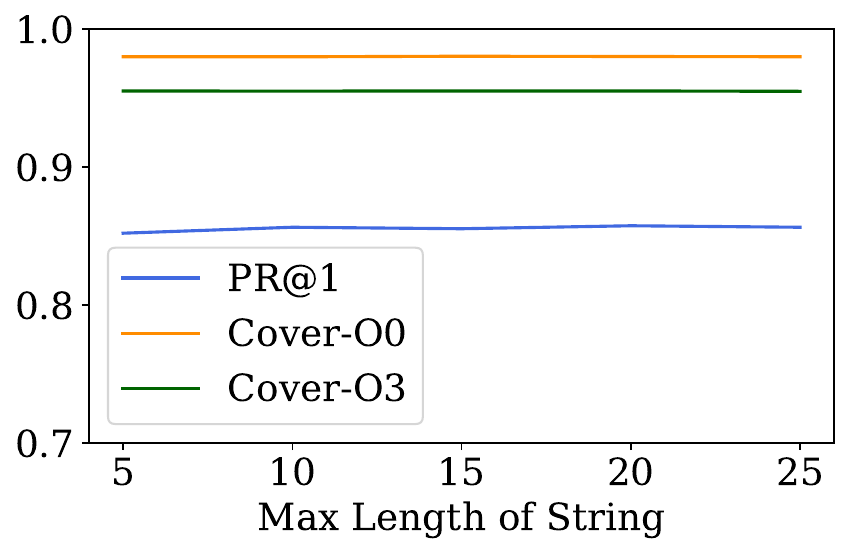}
    \end{subfigure}
    \begin{subfigure}[t]{.2\textwidth}
        \includegraphics[width=\textwidth]{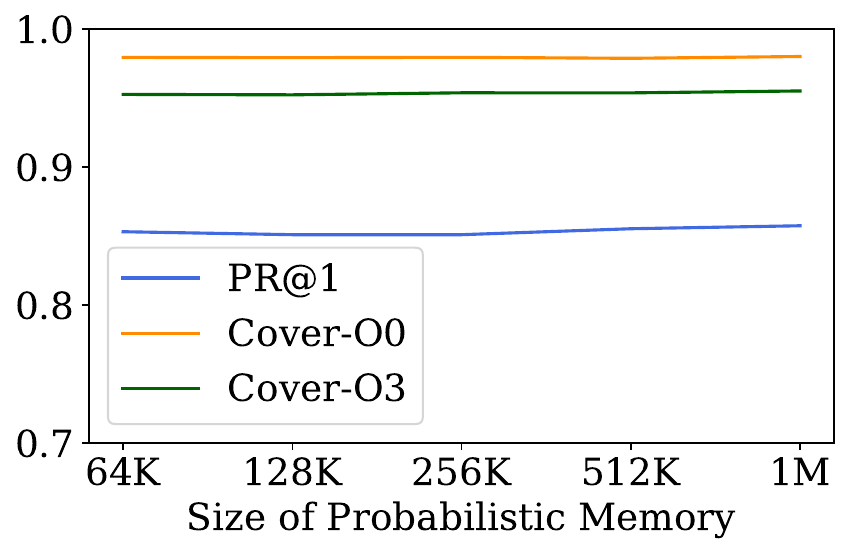}
    \end{subfigure}
    \begin{subfigure}[t]{.2\textwidth}
        \includegraphics[width=\textwidth]{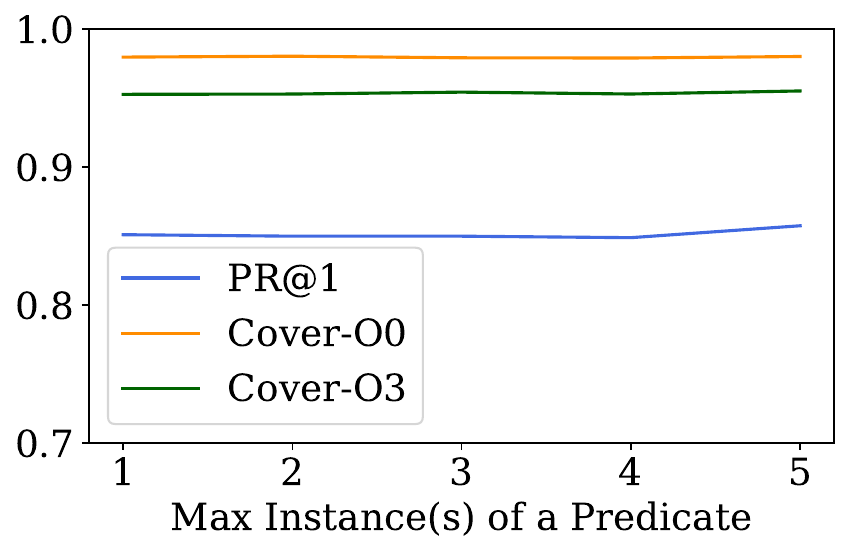}
    \end{subfigure}        
    \caption{Perf. w.r.t. Different Configurations. Each sub-figure presents the experimental results of changing one system configuration. The $x$ axes denote different values of the related system configurations, and the $y$ axes denote the values for  the metrics. The blue lines represent  PR@1, the orange lines  coverage for -O0 programs, and the green lines  coverage for -O3 programs.}
    \label{fig:sys-abl}
\end{figure}

%% file: tab_tex/tab_config.tex
\begin{table}[t]
\footnotesize
    \caption{Perf. w.r.t. Different Configurations. The first two columns list the configurations we alter and the values we use for that configuration. Bold texts are default values for the related configurations. The remaining three columns list the PR@1, coverage for -O0 programs, and coverage for -O3 programs, respectively. The highest value for each column is marked as bold. 
    }
    \centering
    \begin{tabular}{rrccc}
    \toprule
      Configuration & Value & PR@1 & Cover-O0  & Cover-O3 \\ 
    \midrule
    \multirow{5}{*}{Loop-Unroll} & 1    & 0.828       & 0.963      & 0.887  \\
                                & 10    & {\bf 0.862}       & 0.979      & 0.955  \\
                                & {\bf 20}    & 0.857       & 0.980      & 0.955  \\
                                & 30    & 0.856       & 0.980      & 0.954  \\
                                & 40    & 0.853       & 0.980      & {\bf 0.956}  \\
    \midrule
    \multirow{5}{*}{String-Len} & 5    & 0.852       & 0.980      & 0.955  \\
                                & 10    & 0.856       & 0.980      & 0.955  \\
                                & 15    & 0.855       & 0.980      & 0.955  \\
                                & {\bf 20}    & 0.857       & 0.980      & 0.955  \\
                                & 25    & 0.856       & 0.980      & 0.954  \\
    \midrule
    \multirow{5}{*}{Mem-Size} & {\bf 64k}    & 0.853       & 0.979      & 0.952  \\
                                & 128k    & 0.850       & 0.979      & 0.952  \\
                                & 256k    & 0.850       & 0.979      & 0.953  \\
                                & 512k    & 0.855       & 0.978      & 0.953  \\
                                & 1MB    & 0.857       & 0.980      & 0.955  \\
    \midrule
    \multirow{5}{*}{Pred.-Instance} & {\bf 1}    & 0.857       & 0.980      & 0.955  \\
                                & 2    & 0.848       & 0.979      & 0.953  \\
                                & 3    & 0.849       & 0.979      & 0.954  \\
                                & 4    & 0.849       & 0.980      & 0.952  \\
                                & 5    & 0.850       & 0.979      & 0.952  \\
    \midrule
    \multirow{5}{*}{Input-Init} & {\bf Same}    & 0.857       & {\bf 0.981}     & 0.949  \\
                                & Rand1    & 0.845       & 0.980      & 0.950  \\
                                & Rand2    & 0.833       & 0.980      & 0.949  \\
                                & Rand3    & 0.841       & 0.980      & 0.949  \\
                                & Rand4    & 0.841       & 0.980      & 0.950  \\
    \bottomrule
    \end{tabular}
    \label{tab:config}
\end{table}

%% file: fig_tex/fig_attention.tex
\begin{figure*}[t]
    \centering
    \begin{subfigure}[b]{.31\textwidth}
        \begin{minipage}{\textwidth}
        \includegraphics[width=\textwidth]{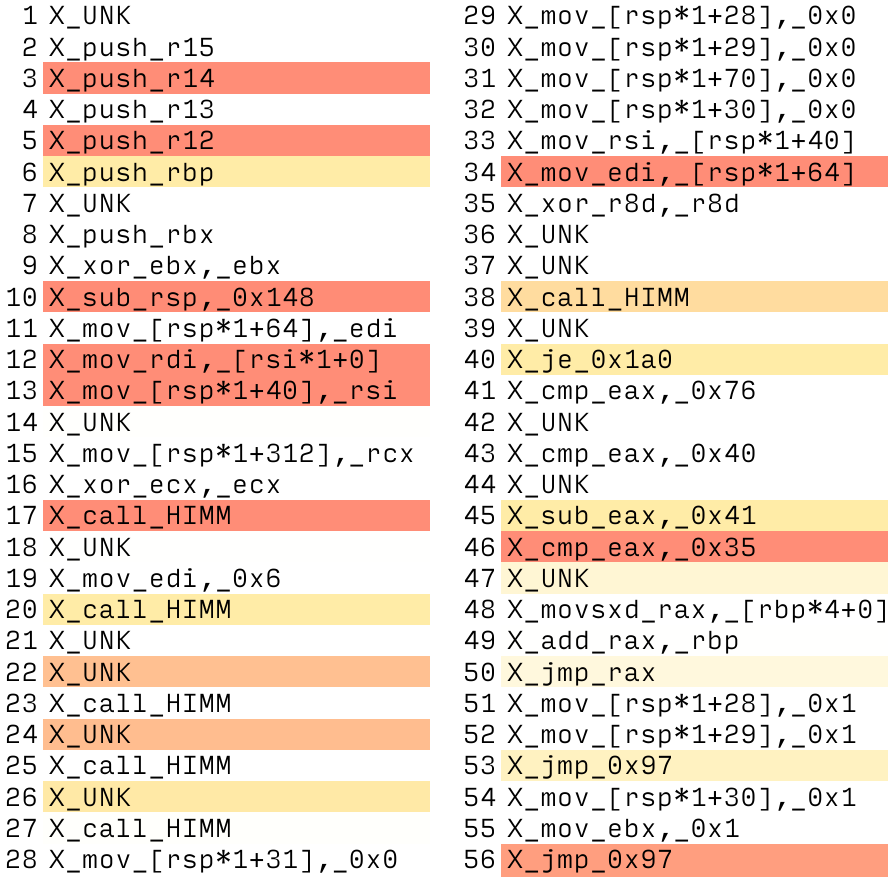}
        \end{minipage}
        \caption{cat O3}
    \end{subfigure}
    \rulesep
    \begin{subfigure}[b]{.31\textwidth}
        \includegraphics[width=\textwidth]{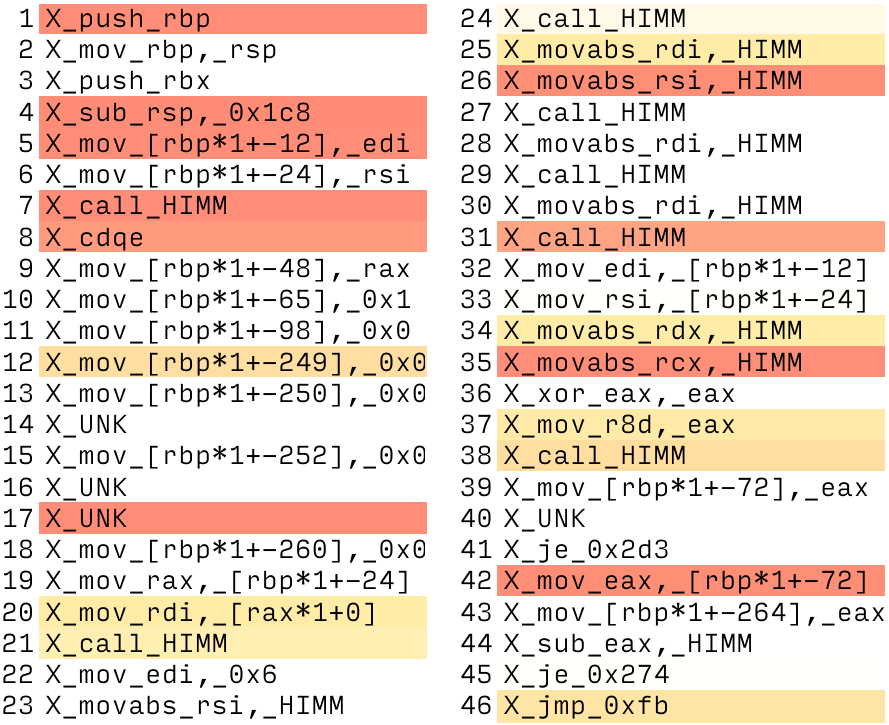}
        \caption{cat O0}
    \end{subfigure}
    \rulesep
    \begin{subfigure}[b]{.31\textwidth}
        \includegraphics[width=\textwidth]{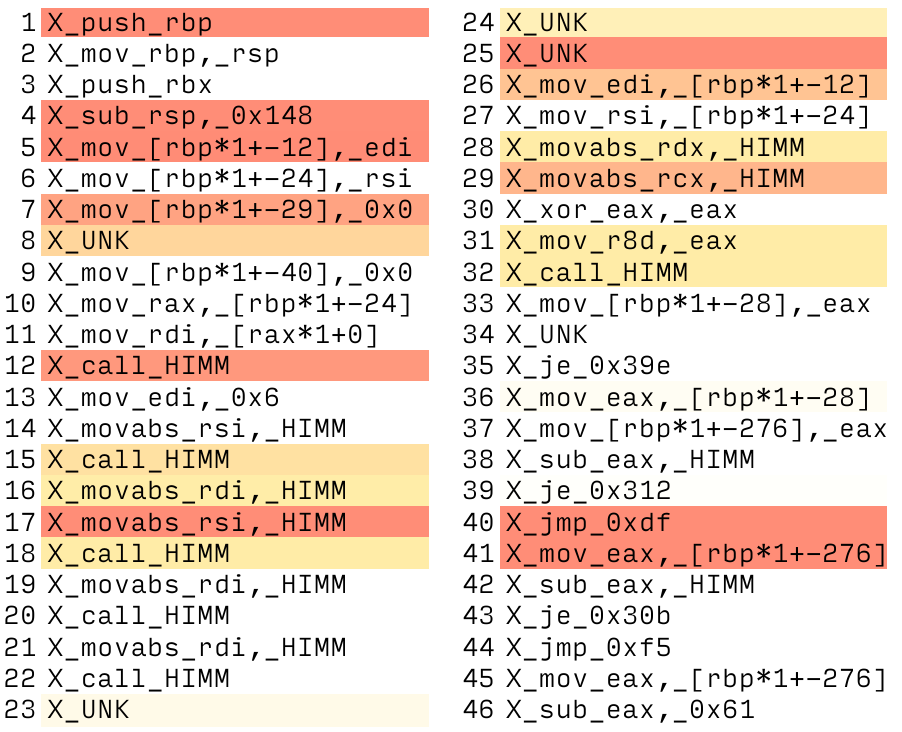}
        \caption{touch O0}
    \end{subfigure}
    \caption{Visualized Attention Layers of SAFE (Darker Color Means Higher Weight)}
    \label{fig:apdx:att}
    \vspace{-5pt}
\end{figure*}

%% file: tab_tex/tab_cve.tex
\begin{table*}[t]
    \caption{Searching for One-Day Vulnerabilities. For each CVE-related function, we query with its optimized version in all the functions from the unoptimized binary. The first column lists the ID of CVE. Columns 2-4 show the project, the number of candidate functions, and the name of the problematic function. The last three columns show the ranking of similarity scores for the ground truth function generated by SAFE, Trex, and \toolname, respectively. An X means we do not find the problematic function within the top 200 matched functions. 
    }
    \centering
    \begin{tabular}{rlllllg}
    \toprule
         \multirow{2}{*}{CVE} & \multirow{2}{*}{Project} & \multirow{2}{*}{\#Func.} & \multirow{2}{*}{Func. Name} & \multicolumn{3}{c}{Result} \\
         \cmidrule(lr){5-7}
         & & & & SAFE & Trex & \cellcolor{white}\toolname \\
     \midrule
    CVE-2017-18018 & Coreutils & 931 & chown & 26 & X & 1\\
    CVE-2007-2452 & Findutils(locate) & 97 & visit\_old\_format & X & 5 & 1\\
    \midrule
    CVE-2021-20311 & \multirow{3}{*}{ImageMagick} & \multirow{3}{*}{1702} & sRGBTransformImage & X & X & 43\\
    CVE-2021-20309 &                             &  & WaveImage & 40 & X & 1\\
    CVE-2019-13308 &                             &  & ComplexImages & 73 & X & 1\\
    \midrule
    CVE-2016-9842 & Zlib & 91 & inflateMark & 34 & 1 & 1\\
    CVE-2021-4044 & Openssl & 2653 & X509\_verify\_cert & X & X & 1\\
    CVE-2021-3711 & Openssl & 2653 & EVP\_PKEY\_decrypt & 10 & X & 1\\
    \bottomrule
    \end{tabular}
    \label{tab:cve}
\end{table*}

%% file: fig_tex/fig_baseline2-2.tex
\begin{figure}[H]
    \centering
    \includegraphics[width=.40\textwidth]{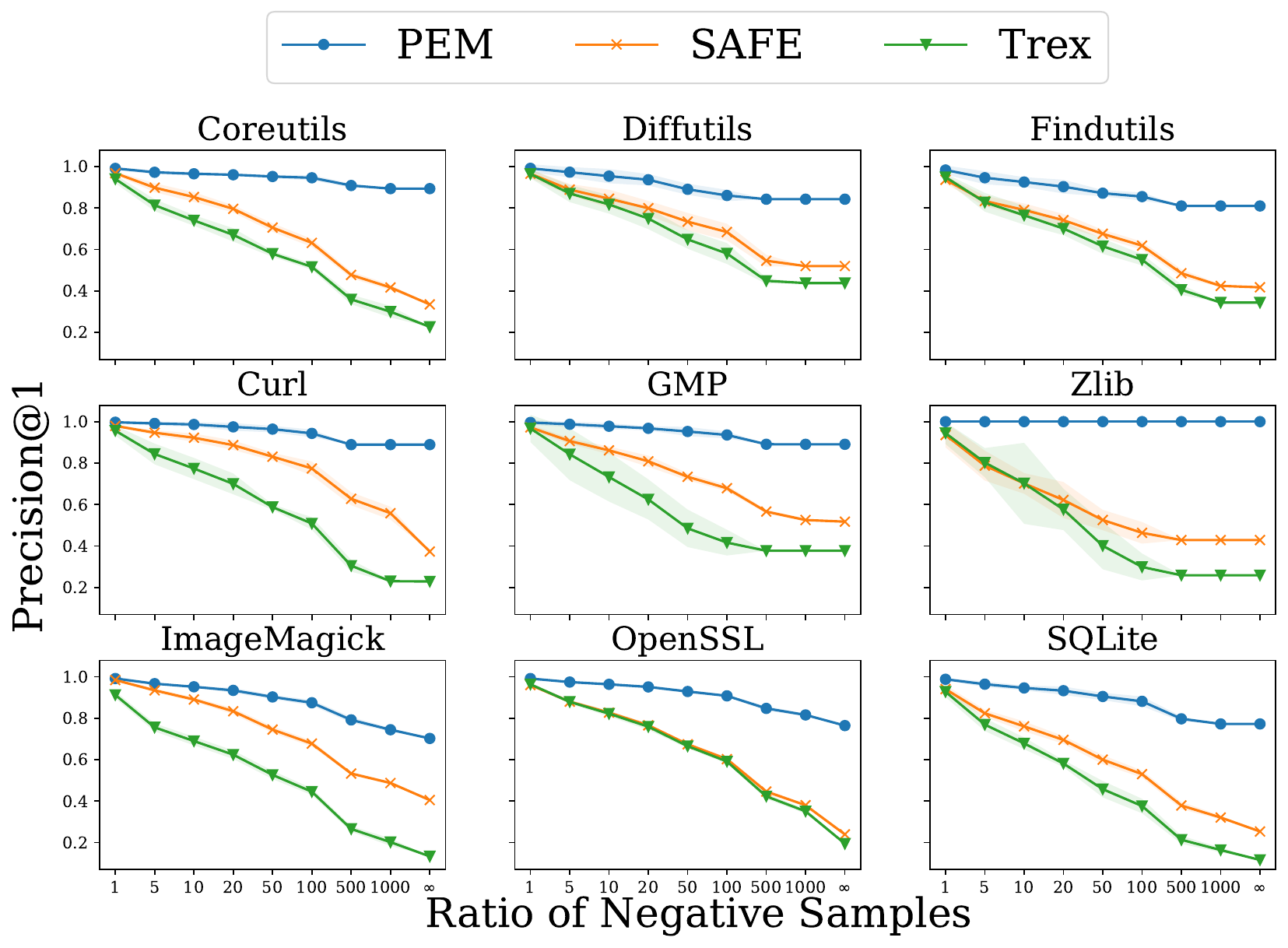}
    
    \caption{
    Variation of PR@1 w.r.t. Negative Sample Ratios. Each figure presents results for one pair of binaries in Dataset-\MakeUppercase{\romannumeral 2}. The $x$ axes denote different ratios of negative samples over positive samples, from 1:1 to $\infty$:1, where $\infty$ is the number of all functions in the binary file. The $y$ axes denote PR@1. Blue, orange, green lines depict performance of \toolname, SAFE, and Trex, respectively.
    }
    \label{fig:baseline2-2}
\end{figure}

%% file: tab_tex/tab_multi_arch.tex
\begin{table}[H]
    \caption{\revise{Cross-architecture Performance on Coreutils. The first column lists the  configurations used to compile the binary files in comparison. \textbf{\textit{X}} and \textbf{\textit{A}} denote X86-64 and AArch 64, respectively. For each row, we use the functions in the binary compiled by the second option to query the similar function in the first binary. The second column lists the precision.}
    }
    \centering
    \revise{
    \begin{tabular}{rccc}
    \toprule
    Pair & PR@1 \\ 
    \midrule
    \textbf{\textit{X}}-O0 \textbf{\textit{X}}-O3 & 89.4\\
    \textbf{\textit{A}}-O0 \textbf{\textit{A}}-O3 & 86.8\\
    \textbf{\textit{X}}-O0 \textbf{\textit{A}}-O3 & 81.2\\
    \textbf{\textit{A}}-O0 \textbf{\textit{X}}-O3 & 81.5\\
    \textbf{\textit{X}}-O0 \textbf{\textit{A}}-O0 & 84.9\\
    \textbf{\textit{X}}-O3 \textbf{\textit{A}}-O3 & 83.9\\
    \bottomrule
    \end{tabular}
    }
    \label{tab:multi-arch}
\end{table}

%% file: tab_tex/tab_statistics.tex
\begin{table}[t]
    \centering
    \caption{\revise{Statistics of Dataset. The first two columns are the names of projects and optimization flags used to compile the binary files. The 3-4 columns are the functions and the number of basic blocks in the related binary files. Note that we remove functions with no more than 3 basic blocks since they are basically wrapper functions.
    }}
    \revise{
    \label{tab:dataset-stats}
    \begin{tabular}{rcrr}
    \toprule
    Name & Opt. & \#Func. & \#B.B.\\
    \midrule
    \multirow{3}{*}{Coreutils-GCC} & O0 & 1943 & 40096 \\
                               & O2 & 1179 & 36318 \\
                               & O3 & 1081 & 42467 \\
   \midrule
    \multirow{3}{*}{Coreutils-Clang} & O0 & 1912 & 52340 \\                               
                               & O2 & 999 & 39099 \\
                               & O3 & 978 & 41816 \\
   \midrule
\multirow{3}{*}{Curl} & O0 & 774 & 17660 \\                               
                               & O2 & 517 & 14514 \\
                               & O3 & 469 & 17102 \\
\midrule
\multirow{3}{*}{ImageMagick} & O0 & 2849 & 102163 \\                               
                               & O2 & 1883 & 103689 \\
                               & O3 & 1817 & 121472 \\
\midrule
\multirow{3}{*}{Diffutils} & O0 & 609 & 12266 \\                               
                               & O2 & 407 & 11540 \\
                               & O3 & 412 & 14298 \\
\midrule
\multirow{3}{*}{Findutils} & O0 & 955 & 18769 \\                               
                               & O2 & 621 & 16212 \\
                               & O3 & 616 & 21943 \\
\midrule
\multirow{3}{*}{OpenSSL} & O0 & 3572 & 68022 \\                               
                               & O2 & 2941 & 58291 \\
                               & O3 & 3054 & 75529 \\
\midrule
\multirow{3}{*}{Zlib} & O0 & 118 & 2779 \\                               
                               & O2 & 106 & 2627 \\
                               & O3 & 98 & 3926 \\
\midrule
\multirow{3}{*}{GMP} & O0 & 605 & 12695 \\                               
                               & O2 & 537 & 14195 \\
                               & O3 & 526 & 16150 \\
\midrule
\multirow{3}{*}{SQLite} & O0 & 1513 & 28224 \\                               
                               & O2 & 1071 & 30636 \\
                               & O3 & 984 & 40171 \\
\midrule
Total & - & {\bf 35146} & {\bf 1077k}\\
\bottomrule
   
    \end{tabular}
    }
\end{table}

%% file: fig_tex/fig_time_eff.tex
\begin{figure}[H]
    \centering
    \includegraphics[width=.29\textwidth]{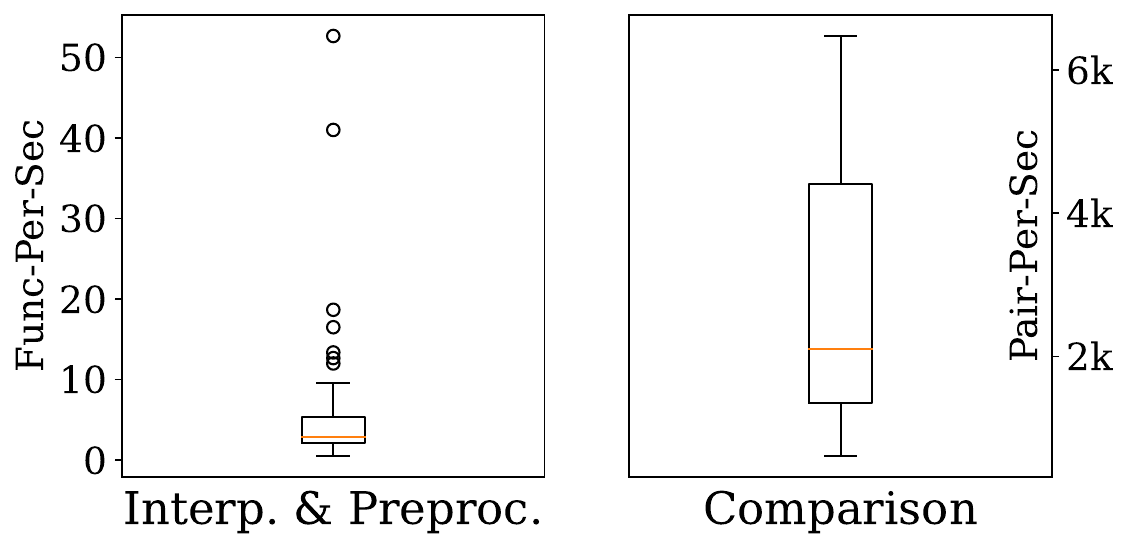}
    \caption{Efficiency 
    The left depicts the throughput for interpretation and preprocessing.
The $y$ axis is the number of functions \toolname can process per second. We can see that \toolname analyzes more than 3 functions per second in most cases.
The right part depicts the throughput for comparison. It illustrates that 
\toolname compares more than 2000 pairs per second in most cases.
    }
    \label{fig:time-eff}
\end{figure}